%% file: sample-sigconf.tex
\documentclass[sigconf]{acmart}
\makeatletter
\renewcommand\@formatdoi[1]{\ignorespaces}
\makeatother

\usepackage{etex}
\renewcommand\footnotetextcopyrightpermission[1]{} 
\setcopyright{rightsretained}
\setcopyright{none}
\renewcommand\footnotetextcopyrightpermission[1]{}
\usepackage{footnote}
\makesavenoteenv{tabular}
\makesavenoteenv{table}

\AtBeginDocument{%
  \providecommand\BibTeX{{%
\normalfont B\kern-0.5em{\scshape i\kern-0.25em b}\kern-0.8em\TeX}}}

\usepackage{caption}
\usepackage{subcaption}
\usepackage{tikz,pgfplots}
\usepackage{listings}
\include{AlgorithmStyle}
\include{mytikz}

\usepackage{array}
\usepackage{makecell}
\newcolumntype{L}[1]{>{\raggedright\let\newline\\\arraybackslash\hspace{0pt}}m{#1}}

\theoremstyle{plain}

\newtheorem{mydef}{Definition} 

\shortauthors{}

\begin{document}

\title{Precise system-wide concatic malware unpacking}

\author{David Korczynski}
\email{david.korczynski@cs.ox.ac.uk}
\affiliation{%
  \institution{Department of Computer Science\\University of Oxford}
}


\begin{abstract}
Run time packing is a common approach malware use to obfuscate their payloads, and automatic unpacking is, therefore, highly relevant. The problem has received much attention, and so far, solutions based on dynamic analysis have been the most successful. Nevertheless, existing solutions lack in several areas, both conceptually and architecturally, because they focus on a limited part of the unpacking problem. These limitations significantly impact their applicability, and current unpackers have, therefore, experienced limited adoption.

In this paper, we introduce a new tool, called Minerva, for effective automatic unpacking of malware samples. Minerva introduces a unified approach to precisely uncover execution waves in a packed malware sample and produce PE files that are well-suited for follow-up static analysis. At the core, Minerva deploys a novel information flow model of system-wide dynamically generated code, precise collection of API calls and a new approach for merging execution waves and API calls. Together, these novelties amplify the generality and precision of automatic unpacking and make the output of Minerva highly usable. We extensively evaluate Minerva against synthetic and real-world malware samples and show that our techniques significantly improve on several aspects compared to previous work.  
\end{abstract}



\maketitle
\thispagestyle{empty}
\section{Introduction}
Conceptually, run time packers encode a binary with obfuscation techniques such as compression and encryption to harden analysis of their code. This hardening significantly increases the effort needed to reverse engineer a given sample, whether manually or automatically, because it requires inverting the anti-analysis techniques used by the packer to understand the full capabilities of the malware. Run time packing is a highly effective anti-analysis technique, and estimates show more than 80\% of malware samples come packed \cite{10.1007/978-3-540-87403-4_6}. The combination of needing to unpack samples before proper analysis is feasible, and that most malware samples come packed makes it desirable to develop approaches that automatically unpack malware.

Techniques and tools for automatic unpacking malware have received a lot of attention in the literature \cite{Dinaburg:2008:EMA:1455770.1455779, Hu:2013:MSM:2535461.2535485, SebastienJosseMalwareDynamicRecompilation,  Kang:2007:RHC:1314389.1314399,  DBLP:conf/malware/Korczynski16, 4413009,  Ugarte-pedrero_sok:deep, DKOR}. Despite this large amount of research, the vast amounts of work rely on the same core principle, the ``write-then-execute'' heuristic. This heuristic deploys the key observation that in order to execute the encrypted code, it first must be decrypted and, therefore, be dynamically generated. The most common approach by previous work is, therefore, to execute a given sample, monitor all memory writes made by the malware, and whenever dynamically written memory executes, the unpacker identifies this memory as decrypted. The tools then dump this specific memory to enable follow-up inspection. 

There are two main limitations to the approach of existing work. First, the ``write-then-execute'' heuristic is not well-suited for packers that perform system-wide unpacking. This is because the heuristic only captures code that is dynamically generated \textit{explicitly} by the malware and not malicious code that is dynamically generated \textit{via} benign code which, unfortunately, is frequently the case in multi-process unpacking. Consequently, existing unpackers are mainly suitable for single-process malware and new approaches to capture system-wide malware unpacking are needed. Second, the primary output of existing work is memory dumps or naively constructed PE files of the dynamically generated code. The output lacks structure, is often an unreasonable over- or under-approximation of the actual malware code, and many obfuscation techniques from the packing process, e.g. obfuscation of external dependencies, remain in the output. As such, the analysis that follows must overcome these obfuscation techniques to enable meaningful analysis of the code. This is a problem because the purpose of unpacking is to facilitate follow-up analysis and not to give any conclusive answer about the malware itself. 

The limitations described above reoccur in existing work, and we argue that an essential reason for this is because existing work widely uses the same set of benchmark applications to validate their solutions. These benchmark applications consist of packers that are out-dated and built more than a decade ago. Consequently, the empirical assessment of novel tools occur with old, and often similar, techniques that do not accurately reflect the challenges posed by modern-day malware packers. To ensure that our novel tools are relevant, we need new benchmark applications that can be used for profiling novel unpackers. These benchmarks must explore corner-cases of modern packing techniques and be easily accessible to anti-malware researchers. 

The goal of this paper is to develop techniques that overcome the limitations of existing work highlighted above. We present a unified approach to precisely unpack malware samples with system-wide execution, dynamically generated code, custom IAT loading and API call obfuscations. The aim is to provide unpacked code that is well-suited for follow-up analysis via manual reverse engineering or off-the-shelf static analysis tools. To this end, Minerva deploys a \textit{combination} of dynamic and static analysis to amplify the effectiveness of automatic unpacking. The novel techniques presented in Minerva rely on information flow, which makes it highly precise and capable of unpacking malware samples in a system-wide context. Minerva models execution waves on a per-process basis and each process with malware execution operate within the context of a single execution wave at any given moment. This provides for a clear wave model and implementation but may result in duplicate content amongst waves, for example, when execution waves use code from an earlier execution wave. 

Minerva takes as input a 32-bit Windows binary and outputs at least one Portable Executable (PE) file per execution wave. This has the benefit of mostly independent PE files but also means the duplicate content of multiple waves will exist in multiple PE files. In order to produce output that is useful for follow-up analysis, Minerva captures how the malware uses external dependencies throughout the entire execution and maps this to each execution wave, resulting in PE files with valid import address tables and patched API calls. Finally, Minerva also performs static analysis to identify relevant malware code within each execution wave. In addition to our unpacker, we also propose a new benchmark suite with applications that combine code-injection techniques, dynamically generated code and obfuscation of external dependencies to overcome the limitations of empirical evaluation in existing work. We demonstrate our unpacker empirically against synthetic and real-world malware samples.  

Our main contributions of this paper are as follows. 
\begin{itemize}
    \item We present a novel approach that combines dynamic and static analysis techniques to unpack malware that executes across the entire system automatically. The approach focuses on precise analysis and outputs unpacked samples that are well-suited for follow-up static analysis. 
    \item We present a new benchmark suite with samples exploring modern-day packing behaviours. To the knowledge of the author, this is the first benchmark suite that comprises synthetic applications aimed at evaluating unpackers. 
    \item We implement the techniques into Minerva and present an extensive empirical evaluation based on synthetic applications and real-world malware samples. 
\end{itemize} 

\section{Background, motivation and overview}
\label{sec:Chapter4BackgroundAndMotivation}
Packing is an umbrella term that refers to a set of various concrete obfuscation techniques and there is no clear definition on the specific obfuscation techniques it encapsulates. This section clarifies the obfuscation techniques we treat in this paper and the limitations of existing work that motivate us. In total, we have compiled six core limitations across two general obfuscation techniques.

\subsection{Dynamically generated code}
The obfuscation technique that is most commonly associated with packing is dynamically generated code. In its simplest terms, dynamically generated code is when an application writes memory at run time and then proceeds to execute this memory. Most often malware does this by containing encrypted code inside its binary image and decrypting this at run time in order to execute it. Existing automated unpackers identify dynamically generated code with the write-then-execute heuristic. This heuristic partitions the malware execution into a set of layers $\mathcal{L}_0$, $\mathcal{L}_1$, \dots $\mathcal{L}_n$ such that each layer constitutes dynamically generated code. Layer $\mathcal{L}_0$ represents the instructions of the binary malware image when first loaded into memory and $\mathcal{L}_{i+1}$ represents the instructions executed on memory written by the instructions in layer $\mathcal{L}_i$. \\

\textbf{Limitation 1.1: the write-then-execute heuristic is unable to capture dynamically generated malicious code \textit{via} benign code.} The strict relationship that instructions of one layer must be dynamically generated explicitly by instructions from a previous layer severely limits the generality of existing work. Malware that uses benign code to dynamically generate its malicious code go unnoticed by this model. The implications of this limitation are substantial for capturing dynamically generated code across multiple processes by way of code-reuse attacks or OS-provided APIs since it is not the instructions of the malicious code that does the writing of memory. Rather, it is benign code that is manipulated by the malware into writing dynamically generated malicious code.  \\

\textbf{Limitation 1.2: existing work unreasonably approximate relevant dynamically generated memory.} Whenever an unpacker observes dynamically generated code it outputs the code for follow-up analysis. To do this, the unpacker must have a definition of what parts of dynamically generated memory are relevant to the unpacked code. This is because not all memory that is dynamically generated, e.g. the stack, is relevant for the unpacked output. However, this step of identifying relevant memory is highly overlooked by previous work. For example, neither Renovo \cite{Kang:2007:RHC:1314389.1314399} nor EtherUnpack \cite{Dinaburg:2008:EMA:1455770.1455779} clearly describe the specific memory they extract during unpacking, and Mutant-X \cite{Hu:2013:MSM:2535461.2535485} dumps the entire memory image of a process when observing dynamically generated code. These are unreasonably imprecise and leave follow-up analysis with the task of identifying a needle in a haystack. \\

\textbf{Limitation 1.3: existing work output raw memory scattered across many memory dumps.} The majority of existing unpackers \cite{Bonfante:2015:CMS:2810103.2813627, Dinaburg:2008:EMA:1455770.1455779, Kang:2007:RHC:1314389.1314399,  Ugarte-pedrero_sok:deep} make little effort to output the unpacked code in a coherent data structure but rather output the unpacked malware in the shape of raw memory dumps. The problem is that when malware dynamically generates code, this may be scattered across several regions, and some of these may also be data-only sections. A precise unpacker should not output incoherent raw memory regions, but rather a suitable data structure that combines these memory regions in an appropriate manner, e.g. re-basing where needed, that enables meaningful follow-up analysis.\\

\subsection{Obfuscating external dependencies}
\label{sec:obfuscatingExternalDependencies}
The way malware interacts with its environment is significant to understanding its malicious activities. We capture this understanding by analysing how the malware uses the OS via  API calls and system calls, and these are, therefore, natural obfuscation targets for malware.  \\

\textbf{Limitation 2.1: existing unpackers fail to accurately correlate  API calls to malicious code.} There is often a large portion of dynamically generated code that must be covered when analysing packed malware. To quickly navigate towards relevant parts, we rely on the malware's use of APIs. However, existing unpackers fail to accurately correlate API calls within a process to the packed code, or even, more generally, attribute whether a given API call was performed by malicious or benign code. Consequently, they report unreasonable estimates of API-usage by the malware.  

In order to circumvent this limitation, the unpacker needs to maintain knowledge of which code belongs to the malware and also be able to identify the instruction responsible for a given API call. From an engineering point of view, most unpackers will be able to augment their systems with solutions to these problems with a modest implementation effort. However, fundamentally, this limitation is guarded by the ability to correctly identify what code belongs to the packed malware, which is closely related to Limitation 1.1 and Limitation 1.2. We identify it here because it is an essential feature in terms of understanding how the unpacked malware code uses external dependencies and something that current unpackers do not support.  For example, when the unpacker from Ugarte et al. is matched with a sample\footnote{sha256 078a122a9401dd47a61369ac769d9e707d9e86bdf7ad91708510b9a4584e8d4} from the Tinba malware family that creates one layer of dynamically generated code before injecting code into the Windows process \texttt{winver.exe}, the unpacker reports that the dynamically generated code performs 1666 API calls from more than 350 different API functions. This result far exceeds the correct count, which is fourteen API calls from ten different API functions. \\

\textbf{Limitation 2.2: output from existing unpackers do not show API-usage when faced with custom API-call resolution.} There is an intricate relationship between dynamically generated code and API call obfuscation. In regular PE binaries, the import address table (IAT) specify the external modules the given application uses. At run time, the operating system linker uses this IAT to load these modules and resolve the addresses of the specific functions the binary imports. However, packed code minimises the IAT to hide how it uses external dependencies, and, instead of using the regular OS linker, the packer deploys a custom linker to resolve its imports. 

\begin{figure}
\centering
\begin{subfigure}[b]{.35\textwidth}
  \includegraphics[width=0.95\linewidth]{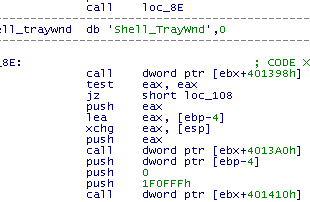}
  \caption{Traditional unpacker, from \cite{Ugarte-pedrero_sok:deep}}
  \label{fig:SoKVSAPIResolution}
\end{subfigure}%

\vspace{\floatsep}

\begin{subfigure}[b]{.35\textwidth}
  \includegraphics[width=0.95\linewidth]{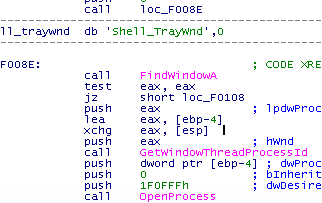}
  \caption{Minerva}
  \label{fig:MinervaVsAPIResolution}
\end{subfigure}
\caption{The output of unpackers when being matched with API calls that are obfuscated with custom API resolution and that branch via a temporal register value.}
\label{fig:SoKvsMinervaUnpackedIATResolution}
\end{figure}

Custom API resolution can happen at any moment(s) during execution and memory dumps taken by existing unpackers are, therefore, susceptible to occur when the malware is yet to resolve its external dependencies. Unfortunately, it is rare that API resolution has occurred the moment dynamically generated code is observed, which is precisely when existing work dumps the memory \cite{Bonfante:2015:CMS:2810103.2813627, Dinaburg:2008:EMA:1455770.1455779, Kang:2007:RHC:1314389.1314399}. Figure \ref{fig:SoKvsMinervaUnpackedIATResolution} shows the differences of matching a traditional unpacker (Figure \ref{fig:SoKVSAPIResolution}) with a sample that has custom API resolution and matching the same sample with an unpacker that accurately captures API-usage in unpacked code (Figure \ref{fig:MinervaVsAPIResolution}), in this case, a result of Minerva. It is clear that without knowledge of the API calls it is hopeless to determine the activities of the code, whereas it is clear from the Minerva-generated code. \\

\begin{figure}
\centering
\begin{subfigure}{.35\textwidth}
  \includegraphics[width=0.95\linewidth]{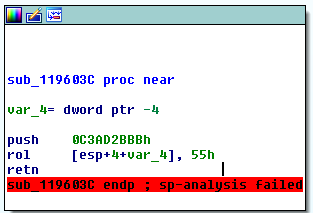}
  \caption{Traditional unpacker}
  \label{fig:PETiteTraditionalUnpacker}
\end{subfigure}%

\vspace{\floatsep}

\begin{subfigure}{.35\textwidth}
  \includegraphics[width=0.95\linewidth]{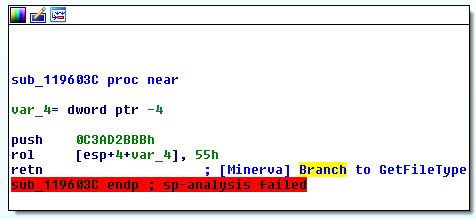}
  \caption{Minerva}
  \label{fig:PETiteMinervaUnpacker}
\end{subfigure}
\caption{The output of unpackers when being matched with an API obfuscation from the PEtite packer.}
\label{fig:PETiteExample}
\end{figure}

\textbf{Limitation 2.3: existing unpackers are unable to identify obfuscated API calls.} Orthogonal to the resolution of API calls, some packed malware samples will go a step further and directly obfuscate the way they call external APIs. In general, there are many ways for malware to obfuscate API calls. In Figure \ref{fig:SoKVSAPIResolution}, we see that the raw calls from Tinba depend on the value of \texttt{EBX}, which in this case contains the base-offset of a custom IAT by the malware. Furthermore, Figure \ref{fig:PETiteExample} shows an example from an application packed with the PEtite\footnote{\url{https://www.un4seen.com/petite/}} packer where the code calls a Windows API function by pushing a value on top of the stack, rotating that value and then transferring execution via a \texttt{ret} instruction to the rotated value on top of the stack. 

The output of existing unpackers is not capable of resolving obfuscated API calls in the unpacked code. This is a problem because it is much harder and sometimes impossible, to determine the destination of the branch instructions in follow-up analysis than it is for the unpacker. For example, without knowledge about the contents of \texttt{EBX}, the data at the address being read and the process layout, it is impossible to determine the destination of the given branch instructions and if they are API calls. 

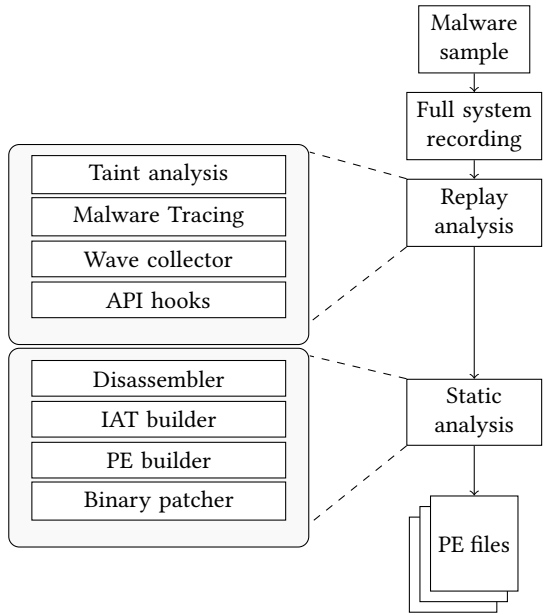
\begin{figure}
    \centering
    \begin{tikzpicture}
    \node(node1) [plain] {Taint analysis};
    \path(node1.east)+(0.2, 0.3) node (invisblenode1) [] {};
    
    \path(node1.south)+(-0.0, -0.3) node (node2)[plain]{Malware Tracing};
    \path(node2.south)+(-0.0, -0.3) node (node3)[plain]{Wave collector};
    \path(node3.south)+(-0.0, -0.3) node (node4)[plain]{API hooks};
    
    \path(node4.east)+(0.2, -0.3) node (invisblenode2) [] {};
    
    \begin{pgfonlayer}{background}
        \path(node1.west |- node2.north)+(-0.3,0.7) node (a) {};
        \path(node1.east |- node2.east)+(+0.3,-1.7) node (c) {};
        \path[fill=gray!05, rounded corners, draw=black] (a) rectangle (c);
    \end{pgfonlayer}
    
    \path(node4.south)+(0.0, -0.8) node (node5)[plain]{Disassembler};
    \path(node5.east)+(0.2, 0.3) node (invisblenode3) [] {};
    \path(node5.south)+(0.0, -0.3) node (node6)[plain]{IAT builder};
    \path(node6.south)+(0.0, -0.3) node (node7)[plain]{PE builder};
    \path(node7.south)+(0.0, -0.3) node (node8)[plain]{Binary patcher};
    \path(node8.east)+(0.2, -0.3) node (invisblenode4) [] {};
    
     \begin{pgfonlayer}{background}
        \path(node5.west |- node6.north)+(-0.3,0.7) node (a) {};
        \path(node6.east |- node6.east)+(+0.3,-1.7) node (c) {};
        \path[fill=gray!05, rounded corners, draw=black] (a) rectangle (c);
    \end{pgfonlayer}   
    
    \path(node1.east)+(2.5, 1.8) node (JDnode)[malplain]{Malware sample};
    
    \path(JDnode.south)+(0.0, -0.7) node (node9)[thickplain]{Full system recording};
    \path(node9.south)+(0.0, -0.7) node (node10)[thickplain]{Replay analysis};
    \path(node10.south)+(0.0, -2.2) node (node11)[thickplain]{Static analysis};

    \path[draw, ->](JDnode.south) -- node[]{} (node9.north);
    \path[draw, ->](node9.south) -- node[]{} (node10.north);
    \path[draw, ->](node10.south) -- node[]{} (node11.north);
    
    \path[draw, dashed](node10.north)+(-0.9, 0.0) -- node[]{} (invisblenode1.east);
    \path[draw, dashed](node10.south)+(-0.9, 0.0) -- node[]{} (invisblenode2.east);
    
    \path[draw, dashed](node11.north)+(-0.9, 0.0) -- node[]{} (invisblenode3.east);
    \path[draw, dashed](node11.south)+(-0.9, 0.0) -- node[]{} (invisblenode4.east);
    
    \path(node11.south)+(0.0, -1.3) node (PEfilesNode)[docs]{PE files};
    
    \path[draw, ->](node11.south)+(0.0, 0.0) -- node[]{} (PEfilesNode.north);

    \end{tikzpicture}
    \caption{Architecture of Minerva's automatic unpacker.}
    \label{fig:MinervaArchitecture}
\end{figure}

\subsection{Solution overview}
The goal of this paper is to develop system-wide, precise and general unpacking techniques. Specifically, our goal is to input a malware binary into our Minerva tool and output PE files that precisely capture the malware code post-decryption and decompression, and also capture how the malware uses external dependencies. The aim is to output PE files that are well-suited for follow-up analysis by off-the-shelf static analysis tools and manual investigation. 

To achieve our goal, we must overcome the limitations highlighted above. First, to overcome the limitations when dealing with dynamically generated code, we need a solution that can (\textit{limitation 1.1}) identify dynamically generated memory across the system; (\textit{limitation 1.2}) extract \textit{precisely} the memory that is relevant to the malware; and (\textit{limitation 1.3}) combine the relevant dynamically generated code into meaningful and related structures. Second, to overcome the limitations against malware that obfuscates external dependencies, the solution must also (\textit{limitation 2.1}) precisely capture the use of API calls \textit{within} the malware code; (\textit{limitation 2.2 and 2.3}) do this in the context of custom API resolution and obfuscated API calls; and, finally, map these observations to the output, so it is readily available for follow-up analysis. 

The solution we come up with, and implement into Minerva, deploys a two-step approach following the architecture shown in Figure \ref{fig:MinervaArchitecture}. First, we use the dynamic analysis in Minerva to precisely extract packed code and the API calls of the malware, and then we use static analysis to construct PE files based on the unpacked code. Specifically, the first step is to capture the malware execution trace using dynamic taint analysis in a similar fashion to Tartarus presented by Korzynski and Yin \cite{DKOR}. Then, we abstract the malware execution trace into execution waves based on information flow analysis such that an execution wave is a process-level construct that represents dynamically generated code in the malware. During the run time analysis, Minerva also ensures precise identification of API calls by the instructions in the malware execution trace. From the first step, we get a set of execution waves consisting of memory dumps, the malware execution trace of each wave and more. The second step Minerva performs is to group related memory within each execution wave using disassembly techniques. Minerva then converts each group of related dumps into a new PE file with a new import address table, and patches API calls based on static analysis and the API calls observed during dynamic analysis. In the following sections, we detail these steps.  

\section{System-wide malware tracing}
A key component of our system is the ability to trace the malware throughout the entire operating system using dynamic taint analysis. We implement the techniques in Tartarus \cite{DKOR} to do this. In order to make this paper self-contained we briefly summarise the idea in this section, however, for complete description of the approach we refer to \cite{DKOR}. 

\subsection{Abstract model of execution environment}
\label{sec:Chapter3AbstractExecutionEnvironment}
We define a formal environment in which we can reason about executions in a sandbox. The model we present is an extension of work from Dinaburg et al. \cite{Dinaburg:2008:EMA:1455770.1455779}. We consider execution at the machine instruction level, and since an instruction can access memory and CPU registers directly we consider a system state as the combination of memory contents and CPU registers. Let $M$ be the set of all memory states and $C$ be the set of all possible CPU register states. We denote all possible instructions as $I$, where each instruction can be considered a machine recognisable combination of opcode and operands stored at a particular place in memory. 

A program $P$ is modelled as a tuple ($M_P$, $\epsilon_P$) where $M_P$ is the memory associated with the program and $\epsilon_P$ is an instruction in $M_P$ which defines the entry point of the program. There are often many programs executing on a system and each of these may communicate with each other through the underlying OS. As such, we model the execution environment $E$ as the underlying OS and the other programs running on the system. 

We define a transition function $\delta_E : I \times M \times C \rightarrow I \times M \times C$ to represent the execution of an instruction in the environment $E$. It defines how execution of an instruction updates the execution state and determines the next instruction to be executed. The trace of instructions obtained by executing program $P$ in execution environment $E$ is then defined to be the ordered set $T(P,E) = (i_0, \dots, i_l)$ where $i_0 = \epsilon_P$ and $\delta_E(i_k, M_k, C_k) = (i_{k+1}, M_{k+1}, C_{k+1})$ for $0 \leq k < l$. We note here that the execution trace does not explicitly capture which instructions are part of the program, with the exception of $i_0$, but rather all the instructions executed on the system including instructions in other processes and the kernel. For any two elements in the execution trace $i_j \in T(P,E)$ and $i_k \in T(P,E)$ we write $i_j < i_k$ if $j < k$, $i_j > i_k$ if $j > k$ and otherwise $i_j = i_k$. We use this to define ordering between the instructions of the sequence. 

\subsection{Malware execution trace}
\label{sec:MalwareExecutionTracing}
We now introduce the concept of malware execution trace. Suppose $P$ is a malware program and $P_A$ is some malware tracer that aims to collect $P$'s execution trace. Malware program $P$ is interested in evading analysis and gain privilege escalation by using code-reuse attacks and code injections. As such, the execution trace of the malware may contain instructions that are not members of program $P$'s memory $M_P$. 

To monitor the malware across the environment, the malware monitor $P_A$ maintains a shadow memory that allows it to label the memory and the CPU registers. This shadow memory is updated for each instruction in the execution trace. Let $S \subseteq M \times C$ be the set of all possible shadow memories. We then define the propagation function $\delta_A : S \times I\rightarrow S$ to be the function that updates the shadow memory when an instruction executes. The list of shadow memories collected by the malware tracer is now defined as the ordered set: $ST_{A}(T(P,E)) = (s_0, \dots, s_l)$ where $\delta_A(s_k, i_k) = s_{k+1}$ for $0 \leq k < l$. 

The job of the malware tracer is to determine for each instruction in the execution trace whether the instruction belongs to the malware or not. To do this, the analyser uses the predicate $\Lambda_A : S \times I \rightarrow \{true, false\}$. The malware execution trace is now given as the sequence of instructions for which $\Lambda_A$ is true and we call $\Lambda_A$ the inclusion predicate. We define the malware execution trace formally as follows: 

\begin{mydef} 
Let $T(P,E)$ be an execution trace and $P_A$ a malware tracer. The malware execution trace is the ordered set $\Pi_A = (m_0, \dots, m_d)$ where: 
\begin{itemize}
\item $\Pi_A$ is a subsequence of $T(P,E)$;
\item $\exists v \, | \, m_j = i_v \wedge \Lambda_A(s_v, i_v)$ \text{ for } $0 \leq j \leq d$.
\end{itemize}
\label{def:real_malware_execution_trace}
\end{mydef}

The above definition says that the malware execution trace is a subsequence (ordering is preserved) of the entire whole-system trace and for each instruction in the malware execution trace there is a corresponding instruction in the whole-system trace for which the inclusion predicate is true. 

The malware execution trace gives us a definition we can use to reason about the properties of malware tracers. In particular, for a given malware tracer it highlights the propagation function, $\delta_A$, and the inclusion predicate, $\Lambda_A$, to be the defining parts. Having constructed our model of malware tracers and identified the key aspects that determine how they collect the execution trace, we now move on to present how Minerva precisely captures system-wide propagation.

\subsection{Tracing the malware execution}
\label{sec:Chapter3OverviewOfMalwareExecutionTracing}
The goal is to capture malware execution throughout the whole system in a precise and general manner. The overall idea is to use dynamic taint analysis to mark the malware under analysis as tainted and then capture its system-wide execution by following how the taint propagates through the system. 

Algorithm \ref{Chapter4:WaveCollection} gives an overview of our approach to capturing the malware execution trace. Assuming the first instruction executed on the system is the entry point of the malware, the first step \textbf{(line 1)} is to taint the memory making up the malware. In particular, we taint the entire malware module, including data and code sections. Next, execution continues until there is no more taint or a user-defined timeout occurs, and for each instruction executed we check if the memory making up the instruction is tainted \textbf{(line 7)}. We include the instruction in the malware execution trace if the instruction is tainted \textbf{(line 8)}. For each instruction in the malware execution trace we taint all the output of the instruction, so as to follow memory generated by the malware that is generated independently of the initial state of the malware memory, as shown by the Update algorithm in Algorithm \ref{alg:UpdateTaint} \textbf{(line 3-5)}.

\section{Information flow execution waves}
\label{sec:DynamicallyGeneratedCode}
Given the malware execution trace, the next step is to partition it into execution waves. The goal of execution waves is to capture dynamically generated malicious code independently of who wrote the code and on the basis that the generated code must originate from the malware. However, we consider execution waves to be more than just a sequence of instructions. The set of execution waves gives an explicit representation of an entire application, including dynamically generated malicious code, and each execution wave may, therefore, include both executable and non-executable data. 

In this section we give a semantics for execution waves (Section \ref{sec:chapter4ExecutionSemantics}) and describe how we collect the waves in practice (Section \ref{sec:chapter4CollectingExecutionWaves}). 

\subsection{Execution wave semantics}
\label{sec:chapter4ExecutionSemantics}
The goal of our execution wave semantics is to clearly define the conversion of a malware sample's execution into waves of dynamically generated malicious code. As such, we describe the waves in relation to an execution trace $T(P,E)$ described in Section \ref{sec:Chapter3AbstractExecutionEnvironment}. 

We partition the malware execution into waves on a process-level basis. We map every instruction in the malware execution trace $i \in \Pi_A$ to a process $P_y$ and a wave within this process $W_x$. We denote $P_yW_x$ to mean wave $x$ within process $y$, and every process with malicious code execution contains a sequence of waves $P_y.\Omega = P_yW_0, \dots, P_yW_n$ with $|P_y.\Omega| \geq 1$. We denote the initial wave in which malware execution begins as $P_{\epsilon}W_{\epsilon}$ and the set $\Phi_{\Pi}$ contains all execution waves for a given malware execution trace $\Pi_A$. For each instruction in the malware execution trace, we first identify the process in which they execute and then the wave they belong to within their respective process. 

Formally, we define an execution wave as follows. 

\begin{mydef} 
An execution wave is a tuple composed of:
\begin{itemize}
    \item A sequence of instructions $\mathcal{I} = i_0, \dots, i_n$ executed in the given wave. We have $i_0$ to be the entry point of the wave;
    \item a shadow memory $\mathcal{S}$, which is a set of ordered pairs $(m_{addr}, m_{byte})$ that contains the tainted memory making up the wave, including both code and data memory;
    \item the tainted writes $\mathcal{T}$ which is a set of ordered pairs $(t_{addr}, t_{byte})$ that holds the tainted memory written by instructions in $P$ since $i_0$, where $P$ is the process of the execution wave.
\end{itemize}
\label{def:execution_wave}
\end{mydef}

Next, we present a set that formalises our requirements for partitioning a complete execution trace into a set of execution waves. The purpose of this definition is to capture every layer of dynamically generated malicious code and not restrict a minimal overlap between the content of each execution wave. In the following, we write for two instructions $i, j$, $i < j$ if $i$ comes before $j$ in the malware execution trace, and vice versa. 

\begin{mydef}
\label{executionWaveDefinition}
Let T(P,E) be an instruction execution trace and $\Pi_A$ the corresponding malware execution trace. The set of execution waves is then given $\Phi_{\Pi} = \{P_0, \dots, P_n\}$ where:

\begin{itemize}
    \item $\forall i \in \Pi_{A} \exists P_x \in \Phi_{\Pi} | i \in P_x.\mathcal{I}.$

    \item For any $P_yW_x$ and $P_yW_z$ in $\Phi_{\Pi}$ where $x < z$ we have that $\forall i_x \in P_yW_x.\mathcal{I}, \forall i_z \in P_yW_z.\mathcal{I} | i_x < i_z$.
    
    This says that there is a strict ordering in the malware execution trace between the instructions of any two waves in a given process $P_y.\Omega$. 

    \item $\forall (m_{addr}, m_{byte}) \in P_wW_{w'}.\mathcal{S} \exists (t_{addr}, t_{byte}) \in P_tW_{t'}.\mathcal{T} \\ | (m_{addr}, m_{byte}) \in P_{\epsilon}W_{\epsilon}.\mathcal{S} \vee (m_{addr}, m_{byte}) = (t_{addr}, t_{byte})$  where $P_wW_{w'} \neq  P_tW_{t'}$ and $\forall i_w \in P_wW_{w'}.\mathcal{I} \exists i_t \in P_tW_{t'} | i_t < i_w$. 
    
    This says that the shadow memory for all execution waves must either exist in the shadow memory of the initial wave or be composed of tainted memory written by a wave that started earlier. 
    
    \item For any wave $P_yW_x \in \Phi_{\Pi}$ we have that $\forall i \in P_yW_x.\mathcal{I} \\ | \exists (m_{addr}, m_{byte}) \in P_yW_x.\mathcal{S} | i[A] = m_{addr}$.
    
    This says the memory of any instruction in each execution wave must be present in the shadow memory of the given wave.
\end{itemize} 
\end{mydef}

An important aspect of Definition \ref{executionWaveDefinition} is that the second bullet enforces a strict ordering between instructions in the set of execution waves for each process. The effect of this is that we preclude instructions from any given execution wave to be used in any other execution wave. The reason we do this is that it creates a clear history of execution wave progress within each process, and it becomes easier to implement since it is only necessary to maintain one execution wave per process. The drawback is that when malware transfers execution to code from an earlier execution wave, we include some content of the earlier execution wave into the current execution wave. In this way, we may end up with waves that overlap in their shadow memory, but, naturally, this can be stripped during post-processing. However, we have found this to be no major issue and that the trade-off works well in practice. However, we leave the door open and encourage future work in other models, e.g. more refined models.

\subsection{Collecting the execution waves}
\label{sec:chapter4CollectingExecutionWaves}

In practice, we only associate one wave with a given process at any given moment. Therefore, to collect the execution waves, it is sufficient to keep track of the current wave in each process. We initially only have one, wave which is the wave inside of the process executing the malicious application. The shadow memory $\mathcal{S}$ of this wave is the malware module when loaded into memory, and the set of tainted writes is initially the empty set, $\mathcal{T} = \emptyset$. We then update the set of tainted writes whenever an instruction writes tainted memory following our Update function shown in Algorithm \ref{alg:UpdateTaint}.

\SetAlgoNoEnd
\begin{algorithm}
 \KwData{(input) Malware sample $B$}
 \KwResult{Logged malware execution waves and malware execution trace $\Pi$.}
 $\mathcal{P} \leftarrow$ init\_taint$(B)$\;
$\mathcal{T}, \mathcal{S} \leftarrow$init\_waves$(B)$ \tcp*{initialise the shadow memories and tainted writes.}
// Full system instrumentation\;
$\mathit{i} \leftarrow first\_instr()$\;
\While{$\mathcal{P} \neq \emptyset$}
{ 
	 // is the instruction tainted?\;
	 \If{$\mathit{i[A]} \in \mathcal{P}$}  
	 {
 		$\Pi \leftarrow \Pi  ^{\wedge} \langle \mathit{i} \rangle$\;
 		
 		\eIf{$i[A] \not \in S_{pid}$}
 		{
 		    \eIf{$i[A] \not \in \mathcal{T}_{pid}$}
 		    {
 		        $S_{pid} \leftarrow S_{pid} \cup (i[A], i[mem])$ \;
 		        $W_{pid} \leftarrow W_{pid} \cup \{i\}$
 		    }
 		    {
 		        $\mathcal{S}_{pid}$, $\mathcal{T}_{pid}$, $W_{pid} \leftarrow$ dump\_wave() \;
 		    }
 		}
 		{
 		    \eIf{$i[A] \in \mathcal{T}_{pid} \wedge S_{pid}[i[A]] \neq i[mem]$}
 		    {
 		        $\mathcal{S}_{pid}$, $\mathcal{T}_{pid}$, $W_{pid} \leftarrow$ dump\_wave() \;
 		    }
 		    {
 		        $W_{pid} \leftarrow W_{pid} \cup \{i\}$
 		    }
 		}
	 }
    $i, \mathcal{P}, \mathcal{T} = update(i, \mathcal{P}, \mathcal{T})\;$
} 
\Return $(\Pi)$
 \caption{Wave collection}
 \label{Chapter4:WaveCollection}
\end{algorithm}

To capture execution waves, we monitor for each process the relationship between the currently executing instruction, the shadow memory and the set of tainted writes following Algorithm \ref{Chapter4:WaveCollection}. Specifically, for every tainted instruction in the malware execution trace, there are four possible cases: 

\begin{enumerate}
\item The address of the instruction is not in the shadow memory and not in the tainted writes \textcolor{blue}{(line 10 Algorithm \ref{Chapter4:WaveCollection})};

\item The address of the instruction is not in the shadow memory but in the tainted writes \textcolor{blue}{(line 13 Algorithm \ref{Chapter4:WaveCollection})};

\item The address of the instruction is in the shadow memory and in the tainted writes but the content of the shadow memory is not similar to current instruction \textcolor{blue}{(line 16 Algorithm \ref{Chapter4:WaveCollection})}; 

\item The address of the instruction is in the shadow memory and in the tainted writes and the content of the shadow memory is equivalent to the memory of the current instruction \textcolor{blue}{(line 18 Algorithm 
\ref{Chapter4:WaveCollection})}.
\end{enumerate}

Case (1) happens in two scenarios. The first case is when tainted memory is transferred across processes via shared memory. For example, if tainted memory is written to memory shared by processes $P_1$ and $P_2$ and the instructions performing the writing is in $P_1$, then the tainted writes will not be in $P_2.\mathcal{T}$ or $P_2.\mathcal{S}$ because we only populate $P_2.\mathcal{T}$ if instructions from $P_2.\mathcal{T}$ are writing to the address space of $P_2$. The second case is when code from the current wave transfers execution to code that is part of an earlier wave. This is because the shadow memory of each wave does not propagate to the proceeding wave, but the memory remains tainted nonetheless. Whenever we observe case (1) we add the memory of the instruction to the shadow memory of the current process and also append the instruction to the sequence of instructions in the current wave. In this case, we update the shadow memory of the current wave with the executing instruction.

In case (4) the current instruction is simply part of the current execution wave, and this is by far the most common case. In this case, we append the instruction to the instruction sequence of the current wave. 
In cases (2) and (3) we consider the current instruction to be the entry point of a new execution wave. Specifically, in case (2) the instruction is dynamically generated in a new memory region and in case (3) the instruction is dynamically generated on top of already existing malware code. 

In the event of a new wave, we log information about the current wave following Algorithm \ref{alg:DumpWave}. First, we log the instructions executed in the current wave, tainted writes and the shadow memory, which includes dumping every page in which there is a tainted write and also dumping the shadow memory. Then we set the shadow memory of the next wave to be the tainted writes of the current wave and set the tainted writes to be the empty set.

\begin{algorithm}
 \KwData{(input)Instruction $i$, memory propagation set $P$, Tainted writes $T$.}
 \KwResult{Next instruction $i_{next}$, memory propagation set $P$, Tainted writes $T$}
  $\mathcal{P} \leftarrow propagate\_taint(i, \mathcal{P})$\;  
  $i_{next} \leftarrow exec\_instr(i)$\;
  \If{$\mathit{i}[A] \in \mathcal{P}$}
	 {
	    		\For{$o \in \mathit{i}[O]$}
 		{
 			$\mathcal{P} \leftarrow \mathcal{P} \cup \{o\}$\;
 		}
	 }
	\For {$w \in \mathit{i}[W]$ } 
    {
    	 \lIf{$w \in \mathcal{P}$}
	    {
            $\mathcal{T}[i.pid] \leftarrow \mathcal{T}[i.pid] \cup \{w\}$
        }
    }
    \Return $i_{next}, \mathcal{P}, \mathcal{T}$
 \caption{Update}
 \label{alg:UpdateTaint}
\end{algorithm}

\begin{algorithm}
 \KwData{(input)Current wave $\mathcal{W}$, shadow memories $\mathcal{S}$, Tainted writes $\mathcal{T}$.}
 \KwResult{Updated $\mathcal{S}$, $\mathcal{T}$, $\mathcal{W}$}
 		        LogInstrs($W_{pid}$)\;
 		        LogTaint($\mathcal{T})$\;
 		        LogShadowMem($\mathcal{S}$)\;
                $\mathcal{S}_{pid} \leftarrow \mathcal{T}_{pid}$\;
                $\mathcal{T}_{pid} \leftarrow \emptyset$\;
                $W_{pid} = \emptyset \cup \{i\}$\;
	 
   \Return $\mathcal{S}_{pid}, \mathcal{T}_{pid}, W_{pid}$\; 
 \caption{dump\_wave}
 \label{alg:DumpWave}
\end{algorithm}

The execution waves capture dynamically generated code independent of who wrote the code including dynamically generated \textit{malicious} code via benign code. We achieve this generality because the shadow table is composed of tainted memory and tainted memory propagates through both benign and malicious instructions. Since the tainted code originates from the malware itself, it is dynamically generated \textit{malicious} code. This property distinguishes our technique from previous work and allows it to be more general without losing precision. 

The output from collecting the execution waves is the sequence of waves executed during the malware execution. For each execution wave, we have memory dumps of the tainted memory during its execution and the list of instructions that belong to each wave. As such, we have an explicit representation of each instruction in the malware execution in the form of its raw bytes, and we also have memory dumps of any non-executed malicious (tainted) memory. All of this information will then be used to reconstruct PE files that are effective for follow-up static analysis.  

\section{Precise dependency capture}
\label{sec:ObfuscatedLibraryCalls}
In order for the PE files to be useful for follow-up static analysis, they must show how the malware uses external dependencies. As described in Section \ref{sec:obfuscatingExternalDependencies}, we must consider custom API call resolution and obfuscated API calls. To this end, we capture the destination of every branch instruction in the malware execution trace and check if it corresponds to the beginning of a function in an external module.

To collect the addresses of functions in each process with malware execution, we iterate the export table of every module in the given process and capture the address of every function it exports. We put these functions in a per-process map that pairs function addresses with their respective function names. Minerva also comes with the possibility to speed up this process using pre-calculated function offsets for a given DLL. As such, with pre-calculated offsets, we only need to know the base address of a given imported module inside the malware process to compute the absolute addresses of its exported functions. 

To capture the API functions that the malware calls, we obtain the destination of every branching instruction in the malware execution trace. If the branch destination is in the set of functions exported by any of the dynamically loaded modules within the execution trace, it means the malware performs an API call. We log every API call and for some functions the parameters as well. For many functions in the Windows API, the return value is also essential to understand the semantics of the call. To capture the return value and output parameters, we note the return address of the API call on the stack and read the output of the function whenever the return address executes. We also monitor functions like \texttt{LoadLibrary} to update our export table when processes load new modules. 

Our approach to monitoring API calls precisely captures the API calls performed by instructions in the malware execution trace and do not capture API calls performed by benign code inside a process in which the malware executes. Furthermore, because we know the specific malicious instruction for each execution wave, it is trivial to map API calls to execution waves. This precise mapping highly improves the precision of the analysis in comparison to sandboxes that capture API calls globally within a process since many of these calls are irrelevant to the malware (this is particularly true in code injected processes). 

Minerva currently does not take any efforts when malware hides API usage by way of stolen bytes or copying of the Windows code. Furthermore, if malware deploys inlined library code or statically linked libraries, then Minerva will not consider these as external dependencies. This is a limitation we discuss further in Section \ref{Chapter4:limitations}. 

\section{Static reconstruction of execution waves}
After collecting the execution waves and external dependencies, we need to combine these into PE files. For each execution wave we construct a set of PE files based on the content of their respective shadow memory and for each PE file we need three ingredients: (1) The specific memory pages of an execution wave that makes up the PE file; (2) the PE's IAT; (3) the entry point of the file. 

The static analysis component of Minerva performs three main steps. First, it groups related memory dumps of each execution wave, then identifies external dependencies in each of these memory groups and, finally, builds new PE files based on the results of the two previous steps.

\subsection{Merging over-approximated shadow memories}
The output from collecting the execution waves includes, for each execution wave, page-level memory dumps of the shadow memory and tainted writes. Intuitively, it can seem appropriate to convert all of these memory dumps into one large PE file and use this for static analysis. However, we have found this to be imprecise in practice because it is rarely the case that all of the memory dumps are relevant to the malware. On the one hand, the shadow memory is a conservative approximation as we capture some memory that is not executable code, and some memory is a result of over-propagated taint. On the other hand, we do not want to reconstruct PE files purely based on executed memory since this will miss non-executed, yet still, malicious executable code, and also relevant data sections. 

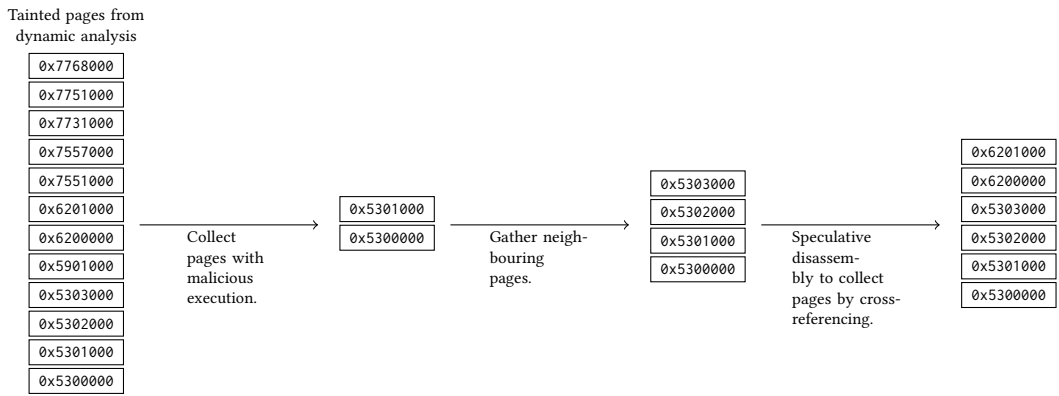
\begin{figure*}[!h]
    \centering
    \begin{tikzpicture}[scale=0.7, every node/.style={transform shape}]
    
    \node(node1) [addressbox] {\texttt{0x7768000}};
    \path(node1.north)+(0.0, 0.5) node (column0text)[textplain]{Tainted pages from dynamic analysis};
    \path(node1.south)+(-0.0, -0.3) node (node2)[addressbox]{\texttt{0x7751000}};
    \path(node2.south)+(-0.0, -0.3) node (node3)[addressbox]{\texttt{0x7731000}};
    \path(node3.south)+(-0.0, -0.3) node (node4)[addressbox]{\texttt{0x7557000}};
    \path(node4.south)+(-0.0, -0.3) node (node5)[addressbox]{\texttt{0x7551000}};
    \path(node5.south)+(-0.0, -0.3) node (node6)[addressbox]{\texttt{0x6201000}};
    \path(node6.south)+(-0.0, -0.3) node (node7)[addressbox]{\texttt{0x6200000}};
    \path(node7.south)+(-0.0, -0.3) node (node8)[addressbox]{\texttt{0x5901000}};
    \path(node8.south)+(-0.0, -0.3) node (node82)[addressbox]{\texttt{0x5303000}};
    \path(node82.south)+(-0.0, -0.3) node (node9)[addressbox]{\texttt{0x5302000}};
    \path(node9.south)+(-0.0, -0.3) node (node10)[addressbox]{\texttt{0x5301000}};
    \path(node10.south)+(-0.0, -0.3) node (node11)[addressbox]{\texttt{0x5300000}};
    
    \path(node6.east)+(3.8, -0.3) node (InvisColumn2){};
    
    \path[draw, ->](node6.east)+(0.3,-0.3) -- node[below, text width=5em]{Collect pages with malicious execution.} (InvisColumn2.west){};
    
    \path(node6.east)+(5.0, -0.0) node (Column2Center)[addressbox]{\texttt{0x5301000}};
    \path(Column2Center.south)+(0.0, -0.3) node (column2node2)[addressbox]{\texttt{0x5300000}};
    
    \path(Column2Center.east)+(3.8, -0.3) node (InvisColumn3){};
    \path[draw, ->](Column2Center.east)+(0.3, -0.3) -- node[below, text width=6em]{Gather neighbouring pages.} (InvisColumn3.west){};

    \path(Column2Center.east)+(5.0, -0.6) node (Column3Center)[addressbox]{\texttt{0x5301000}};
    \path(Column3Center.south)+(0.0, -0.3) node (column3node2)[addressbox]{\texttt{0x5300000}};
    \path(Column3Center.north)+(0.0, +0.3) node (column3node3)[addressbox]{\texttt{0x5302000}};
    \path(column3node3.north)+(0.0, +0.3) node (column3node4)[addressbox]{\texttt{0x5303000}};
    
    \path(Column3Center.east)+(3.8, +0.3) node (InvisColumn4){};
    \path[draw, ->](Column3Center.east)+(0.3, +0.3) -- node[below, text width=7em]{Speculative disassembly to collect pages by cross-referencing.} (InvisColumn4.west){};

    \path(Column3Center.east)+(5.0, +0.6) node (Column4Center)[addressbox]{\texttt{0x5303000}};
    \path(Column4Center.south)+(0.0, -0.3) node (Column41Center)[addressbox]{\texttt{0x5302000}};
    \path(Column41Center.south)+(0.0, -0.3) node (column4node2)[addressbox]{\texttt{0x5301000}};
    \path(column4node2.south)+(0.0, -0.3) node (column4node3)[addressbox]{\texttt{0x5300000}};    
    \path(Column4Center.north)+(0.0, +0.3) node (column4node4)[addressbox]{\texttt{0x6200000}};
    \path(column4node4.north)+(0.0, +0.3) node (column4node5)[addressbox]{\texttt{0x6201000}};
    \end{tikzpicture}
    
    \caption{The process of identifying which tainted pages from dynamic analysis that are relevant when reconstructing unpacked PE files. In the example, the reconstructed PE file has two sections (\texttt{0x5300000}-\texttt{0x5303000} and \texttt{0x6200000}-\texttt{0x6201000}).}
    \label{fig:SelectingTaintedPagesForPE}
    
\end{figure*}

To avoid this imprecision, we divide the page-level memory dumps from the dynamic analysis into smaller groups, such that the pages of each group are related and no page in a given partition relates to any other partition. The goal with this is to capture the parts of the malware that are self-contained and represent the application timelessly. To this end, we create a PE file with multiple sections for each partition. Figure \ref{fig:SelectingTaintedPagesForPE} shows an example of how we select the specific tainted pages that are relevant for the unpacked malware from a set of tainted pages output by Minerva's dynamic analysis component. 

The first step is to identify the tainted pages with malicious code execution. To do this, we first iterate the sequence of instructions executed in a given execution wave and collect all pages that hold instructions from this sequence. Following this, we iteratively collect neighbouring pages until there are no more neighbouring pages and the result is a set of page-level intervals where some pages hold executed code, and other pages neighbour up to these. This corresponds to the first two steps in Figure \ref{fig:SelectingTaintedPagesForPE}. 

Following this, we identify pages in the shadow memory that relate to each interval. To construct self-contained PE files, we capture data-dependencies and control-dependencies to other pages in the shadow memory for each interval. We do this by performing speculative disassembly on each memory dump to capture cross-references to other memory dumps. This step gives us cross-references for each interval, and we then iteratively merge related intervals such that no interval will have cross-references to other intervals. Following this approach, we end up with a set of groups of memory dumps, and we create a PE file for each of these groups. In the example in Figure \ref{fig:SelectingTaintedPagesForPE} we end up with one group consisting of two intervals and will, therefore, create one PE file. 

\subsection{Dependency reconstruction}
To reconstruct external dependencies in our PE files, we need to rebuild the IAT of the binary and patch instructions to rely on this new IAT.

To construct the IAT, we first identify API calls made by instructions belonging to the pages of each memory group. We identify these by matching the API hooks collected during dynamic analysis to instructions of the respective code wave and the pages of the given memory group. We include each unique API function in the IAT of the reconstructed PE file. 

Although we know which instructions branch to external APIs from the malware execution trace, the branch destinations may not be visible from the memory dumps themselves. The final step in constructing PE files is, therefore, to map the instructions that perform API calls to our newly generated IAT by patching them on the binary level. Unfortunately, binary patching is not an easy task since some instructions may require for us to rearrange the instructions in the binary, and this may subsequently break it. In practice, we patch branch instructions that are 6 bytes long, e.g. \texttt{call [0xdeadbeef]}, because we can do this without rearranging instructions. We do not patch instructions that are less than 6 bytes, e.g. \texttt{call eax}. However, we still keep the cross-references so they can be used in more abstract representations in a follow-up analysis. 

\subsection{Final PE construction}
In order to construct the final PE file, we need to know the entry point and the PE sections to put in the file. To identify the entry point, we go through the instruction sequence of the given wave and identify the first instruction in the range of each memory dump group. In order to construct the PE sections, we rely on the memory intervals that we end up with in each memory group. For example, in our example from Figure \ref{fig:SelectingTaintedPagesForPE} we end up with one group and two intervals ([\texttt{0x5300000}-\texttt{0x5303000}], [\texttt{0x6200000}-\texttt{0x6201000}]). We make each of these intervals into an individual section of the PE file and place the newly generated IAT in-between the PE header and these sections. The reason we make each of them into individual sections is to avoid rebasing each interval. The pages we dump from virtual memory are placed at various locations, and each of them must keep this virtual address in the PE file. As such, the \texttt{PointerToRawData} and \texttt{VirtualAddress} values in each section header will be significantly different, and the \texttt{VirtualAddress} points to the base address of each interval as it was when dumped from virtual memory (\texttt{0x5300000} and \texttt{0x6200000} in the example in Figure \ref{fig:SelectingTaintedPagesForPE}). 

\section{Evaluation}
\label{sec:Chapter4Evaluation}
Having presented the core techniques of Minerva, we now move on to evaluate Minerva using multiple benchmarks with respect to the following research questions:
\begin{enumerate}
    \item Does Minerva precisely capture dynamically generated code and the malware's API calls?
    \item Does Minerva improve results over previous work?
    \item Is Minerva relevant for common malware analysis tasks? 
\end{enumerate}

To facilitate the research questions above, we gather four sets of benchmark applications comprising synthetic applications as well as real-world malware applications:
\begin{enumerate}
    \item \textbf{Benchmark \#1 : Ground truth data set.} We develop a new benchmark suite that combines the use of dynamically generated code, code injection and obfuscation of external dependencies. In total, we have developed nine different applications, and they are all described in Table \ref{tab:Chapter4DataSetADynamic}. 
    
    The applications in the benchmark suite represent many of the challenges posed by real-world packers. To the knowledge of the authors, this is the first dedicated benchmark suite for challenging the attributes of unpackers where none of the samples relies on packers developed by third-party teams. The benefit of this benchmark suite is that each sample poses specific challenges that are clearly defined, the applications are easy to understand, and we have the complete source code of each example. As such, it becomes much more accessible to determine if an unpacker is successful because there is no need to reverse engineer large amounts of binary code. 

    \item \textbf{Benchmark \#2 : Selected malware samples.} The second data set corresponds to several malware samples from the families CryptoWall, Tinba, Gapz and Ramnit. These samples perform many of the obfuscation techniques that Minerva aims to overcome, such as code injection combined with dynamically generated code and custom API resolution. 

    \item \textbf{Benchmark \#3 : Packed synthetic samples.} We have taken a set of synthetic samples and packed them with well-known packers. In these applications, we know the applications' behaviours before packing because we design the applications; however, we do not know the exact changes the packers make on the code and, therefore, do not have ground truth about the packed applications.
    
    \item \textbf{Benchmark \#4 : Real-world malware samples.} This set comprises 119 malware samples from the real-world malware families listed in Table \ref{tab:Bench4MalwareFamilies}. We collected seven samples from each family to maintain a balanced data set, and the samples were collected from VirusTotal. In order to ensure the samples are indeed benign, we required each sample to be detected by at least 15 anti-malware vendors. Furthermore, in order to ensure certainty that the samples belong to their respective families, we required at least two vendors to label them in the same family. On average each sample had 52 anti-malware vendors report it as malicious and a median of 54. We recorded each of these samples for 25 seconds and set a max replay time of 120 minutes. 
\end{enumerate}  

\begin{table}{}
    \centering
    
    \begin{tabular}{l|c|c|c}

    Artemis & CTBLocker &  Cerber & CoinMiner \\
    \hline
        CosmicDuke & Emotet & Kovter & Madangel \\
    \hline
        Mira & Natas & Nymaim & Pony \\
    \hline
        Shifu & Simda & TinyBanker & Urausy \\
    \hline
        Zbot & &&  \\
    \end{tabular}
    \caption{The malware families in Benchmark set \#4. We collected a total of seven samples from each family.}
    \label{tab:Bench4MalwareFamilies}
\end{table}

In order to assess the techniques of Minerva, we must make a fair and meaningful comparison to existing work. One approach is to compare Minerva to recently proposed unpackers like Codisasm \cite{Bonfante:2015:CMS:2810103.2813627} or Aranchino \cite{10.1007/978-3-319-60876-1_4}. However, we already showed in \cite{DKOR} that Codisasm is very limited due to its implementation in PIN, and Aranchino is also developed on top of PIN with no additional effort for analysing system-wide malware. Instead, we compare Minerva to the unpacker by Ugarte et al. \cite{Ugarte-pedrero_sok:deep}. 

Ugarte et al. \cite{Ugarte-pedrero_sok:deep} propose a malware unpacker that is capable of analysing multi-process malware is implemented on top of QEMU. The unpacker supports multi-process unpacking by monitor various system calls and also develop techniques for capturing memory mappings. The tool they present is only available as a web service\footnote{www.packerinspector.com}, which forces us to treat their system as a black box. Furthermore, they do not mention which OS they support in their work; however, from experimenting with the service, we conclude the analysis environment is Windows XP. We determined this because the web service responds with \textit{``Error - The sample did not start executing.''} when faced with applications compiled for Windows 7 and later, but runs normally with Windows XP applications. As such, we wrote the samples in our data set to make sure they all execute correctly on both Windows XP and Windows 7. We will refer to the unpacker by Ugarte et al. \cite{Ugarte-pedrero_sok:deep} as PackerInspector.

\begin{table}{}
\footnotesize

\begin{tabular}{| l | L{7.5CM} |}
\hline
\textbf{ID} & \textbf{Description}. \\
\hline
D1 & Dynamically generates code and uses custom IAT resolution to resolve \texttt{GetModuleHandle}, \texttt{GetProcAddress} and \texttt{ExitProcess} and exits.  \\
\hline
D2 & Dynamically generates code and uses custom IAT resolution to resolve \texttt{GetModuleHandle}, \texttt{GetProcAddress} and \texttt{MessageBoxA} and then displays a message box.\\
\hline
D3 & Dynamically generates code that further dynamically generates code and then uses custom IAT resolution to resolve \texttt{GetModuleHandle}, \texttt{GetProcAddress} and \texttt{MessageBoxA} and then displays a message box.\\
\hline
D4 & Dynamically generates code that further dynamically generates code and then uses custom IAT resolution to resolve \texttt{GetModuleHandle}, \texttt{GetProcAddress} and \texttt{ExitProcess} and then exits.\\
\hline
C1 & Opens the Windows process \texttt{explorer.exe} using \texttt{OpenProcess, WriteProcessMemory} and \texttt{CreateRemoteThread}, then inside the target process dynamically resolves the address of \texttt{GetModuleHandle}, \texttt{GetProcAddress} and \texttt{ExitProcess}, and calls each of them to exit.\\
\hline
C2 & Opens the Windows process \texttt{explorer.exe} using \texttt{OpenProcess, WriteProcessMemory} and \texttt{QueueUserAPC}, then inside the target process dynamically resolves the address of \texttt{GetModuleHandle}, \texttt{GetProcAddress} and \texttt{ExitProcess}, and calls each of them to exit.\\ 
\hline
C4 & Uses the PowerLoaderEx injection that relies on a global memory buffer and code-reuse attacks to hijack execution of \texttt{explorer.exe}. Inside \texttt{explorer.exe} code-reuse attacks transfers execution to shellcode that calls \texttt{LoadLibraryA}.\\
\hline
C5 & Uses the Atombombing injection techniques that relies on the global atom tables to execute within \texttt{explorer.exe}. Inside \texttt{explorer.exe} it uses code-reuses attack to execute a piece of shellcode that launches \texttt{calc.exe}.\\
\hline
M1 & Injects code into \texttt{explorer.exe} similarly to A1, then inside the target process dynamically generates code that then dynamically resolves the address of \texttt{GetModuleHandle}, \texttt{GetProcAddress} and \texttt{ExitProcess}, and calls each of them to exit.\\
\hline
\end{tabular}

\caption{Description of the samples in data set \#1 and how they perform code injection.}
\label{tab:Chapter4DataSetADynamic}
\end{table}

\subsection{Implementation}
Minerva is built on top of PANDA \cite{Dolan-Gavitt:2015:RRE:2843859.2843867}, which is a dynamic analysis framework based on full system emulation and utilises a record-and-replay infrastructure. All of the code on top of PANDA is built in C/C++ and the majority of our tools that process the output of the sandbox are in Python. Most of the code in Minerva's dynamic analysis is on top of PANDA; however, we have had to modify the main taint analysis plugin that comes with PANDA to be less resource intensive. Specifically, PANDA's \texttt{taint2} plugin can quickly use 40+ GB of memory, and to limit this, we removed support for taint-labels and made some data structures more simplistic. 

\subsection{Experimental set up}
\label{sec:Chapter5ExperimentalSetup}
We conduct all of our Minerva experiments on a 4-core Intel-7 CPU with 4.2 GHz and a Windows 7, 32-bit guest architecture. The guest is in a closed network and connected to another virtual machine that performs network simulation using INetSim\cite{InetSim}. As such, malware samples that connect back to some CC server will be able to resolve DNS names, connect to every IP and also receive content. However, the content itself is the default data provided by INetsim. 

We executed the applications on the guest machine with a local admin account, and User Account Control (UAC) enabled. We perform no user stimulation during the analysis, and there were no applications apart from the generic Windows processes running in the guest machine itself. 

\subsection{Empirical evaluation of correctness}
\label{sec:Chapter5EmpiricalEvaluationOfCorrectness}
In our first experiment, we match Minerva and PackerInspector with the ground-truth samples in benchmark set \#1. For each of the samples, we capture the number of execution waves, the number of processes involved in the execution, the number of API calls observed from the last wave of each sample and the number of functions in the IAT of the unpacker's output. We match the results from the output of Minerva and PackerInspector with our ground truth data and Table \ref{tab:ground_truth_evaluation} shows our results.

For the samples that execute in a single process, both Minerva and PackerInspector capture the number of processes and waves accurately. In two of these four samples, Minerva captures the five expected API calls accurately, and in the other two Minerva captures slightly more than the expected number. PackerInspector, however, attributes about 200x more API calls than the expected number to the final wave of the execution. Furthermore, Minerva builds the IAT for the two samples accurately and a slightly larger IAT for the other two. PackerInspector is unable to produce any output with an IAT, and there is no sign of API usage in the output of PackerInspector. The reason Minerva captures slightly more API calls than expected is that the compiler, naturally, adds various function calls around the source code. PackerInspector successfully identifies the correct number of execution waves but fails to attribute API calls accurately to unpacked code and also fails to produce any output with an IAT. Minerva, however, succeeds at both. 

For the samples that perform multi-process execution, we observe that Minerva captures all processes, execution waves, API calls, and rebuilds PE files with the expected IAT. The reason Minerva does not capture slightly more API calls than the expected amount in these samples is that the final wave occurs within an injected process and does not contain the added functions from the compiler. Surprisingly, PackerInspector fails to detect multi-process execution in any of the samples, and we suspect this is because PackerInspector only monitors for multi-process execution via memory mapped files which none of the samples uses. From our multi-process samples, we observe the limitations of the original write-then-execute heuristic, in that it is unable to handle system-wide unpacking in a general and precise manner. However, the novel techniques introduced by Minerva are successful at this.  

When matched with our ground-truth samples it is clear that PackerInspector over-approximates the API usage of the applications, is unable to output unpacked code that shows API usage when faced with obfuscations of external dependencies and under-approximates the system-wide malware execution. These observations verify our hypothesis that state-of-the-art unpackers are unable to deal with many challenges faced by system-wide packing and that the techniques in Minerva overcome these limitations. 

\begin{table}{}
    \centering
     \footnotesize
    \begin{tabular}{l|c|c|c|c}
    
        & \multicolumn{4}{c}{\textbf{Precision}} \\
        & \multicolumn{4}{c}{(Ground Truth, Minerva, PackerInspector)} \\
         Sample & \#Procs & \#Waves & \#API calls in final wave &\#IAT size\\
         \hline
         \textbf{(1)} D1  & 1,1,1 & 2,2,2 & 5,8,1007 & 3,5,0 \\
         \textbf{(1)} D2  & 1,1,1 & 2,2,2 & 5,5,1007 & 3,3,0 \\
         \textbf{(1)} D3  & 1,1,1 & 3,3,3 & 5,5,1006 &  3,3,0\\
         \textbf{(1)} D4 & 1,1,1 & 3,3,3 & 5,10,1007 & 3,6,0 \\
            \hline
        \textbf{(1)} C1 & 2,2,1 & 2,2,1 & 6,6,$\dagger$ & 3,3,$\dagger$ \\    
         \textbf{(1)} C2 & 2,2,1 & 2,2,1 & 6,6,$\dagger$ & 3,3,$\dagger$ \\
         \textbf{(1)} C4 & 2,2,1 & 2,2,1 & 1,1,$\dagger$ &  1,1,$\dagger$\\
         \textbf{(1)} C5 & 2,2,1 & 2,2,1 & 5,5,$\dagger$ &  3,3,$\dagger$ \\
         \hline
         \textbf{(1)} M1 & 2,2,1 & 3,3,1 & 6,6,$\dagger$ &  3,3,$\dagger$ \\
    \end{tabular}
    \centering
    \caption{The evaluation results from matching Minerva and PackerInspector with the ground-truth samples of data set \#1. $\dagger$ means not available because PackerInspector failed to reach the last wave.}
    \label{tab:ground_truth_evaluation}
\end{table}

\subsection{Empirical evaluation against selected malware}
\label{sec:Chapter5EmpirivalEvaluationAgainstSelectedMalware}
In our second experiment, we match Minerva and PackerInspector with the malware samples in data set \#2. The goal of this experiment is twofold. First, we aim to measure how each unpacker captures system-wide unpacking based on the number of processes and waves they identify. Second, we aim to measure the difference in the total amount and the unique amount of API calls observed in each  malware execution. 

We only have access to PackerInspector via their web interface, and in this experiment, it is important to highlight the limitations of this. The samples we analyse with Minerva are executed in Windows 7, and the samples we analyse with PackerInspector presumably execute in Windows XP. In addition to this, we do not know the state of the execution environment that PackerInspector uses to execute the malware samples, such as the processes executing on the system, the network connection, the privilege-level of the malware, the security settings in the guest system, and so on. This adds a level of uncertainty to the results we present in this section since we are not conducting an isolated comparison of techniques as the execution environments of the malware are, likely, significantly different. This is particularly relevant when dealing with the samples from data set \#2 because malware samples are complex applications that are sensitive to their execution environment and small changes in the environment can have a substantial impact on the malware execution. However, we still feel it is appropriate to report the results as they give certain insights into the differences in our approaches. The results of our experiment are shown in Table \ref{tab:malware_evaluation}.  

In terms of multi-process monitoring, Minerva captures more injections than PackerInspector in eight samples and fewer injections in three samples. PackerInspector misses multi-process executions in all Tinba and Gapz samples. For the Tinba samples, PackerInspector only catches the first multi-process execution which occurs into the benign Windows process \texttt{winver.exe}, a process that is also started by each sample. PackerInspector misses all multi-process executions within the Gapz malware sample, and we believe there are two possible explanations for this. First, because PackerInspector exits prematurely, and second because Gapz uses the PowerLoader injection which does not rely on any of the API hooks used by PackerInspector to catch multi-process unpacking. 

In one CryptoWall sample, Minerva finds one more multi-process propagation than PackerInspector, and in another sample, Minerva finds one less than PackerInspector. In both samples, both unpackers find two injections into \texttt{svchost.exe} and in both cases, PackerInspector also finds injections into \texttt{vssadmin.exe}. Minerva also finds an injection into \texttt{vssadmin.exe} in the sample with five process executions. However, we have found that these \texttt{vssadmin.exe} propagations correspond to false positives. In particular, we found no injections into \texttt{vssadmin.exe} but rather that the samples execute the following command \texttt{``vssadmin.exe Delete Shadow /All /Quiet''} using the \texttt{WinExec} call. We believe this call results in the memory flowing into the \texttt{vssadmin.exe} and is, effectively, the reason the unpackers identify execution in \texttt{vssadmin.exe}. 

In two Ramnit samples, Minerva finds fewer injections than PackerInspector. In one of these samples, PackerInspector reports an additional injection into \texttt{IEXPLORE.exe} that is not identified by Minerva. We analysed the sample ourselves and found that the sample only injects into two \texttt{IEXPLORE.exe} processes if it fails to inject into two \texttt{svchost.exe} processes, which we observed by both Minerva and PackerInspector. However, researchers from Symantec \cite{SymantecRamnit} report that Ramnit also drops a file that will be loaded by each new instance of \texttt{IEXPLORE.exe}, which may be an explanation for why PackerInspector observes such an injection. However, we think it's most likely a result of over-approximation in PackerInspector as it would require the \texttt{IEXPLORE.exe} process to be launched on the system. In the other sample, PackerInspector finds an additional injection into a file with a random name which Minerva does not. Minerva, however, observes the creation of this file but does not see it execute. In the remaining Ramnit sample, Minerva captures seven more injections than PackerInspector, and both Minerva and PackerInspector identifies four injections into \texttt{IEXPLORE.exe} in this sample. Based on follow-up analysis, we determine six of the additional injections Minerva finds are true positives and one is a false positive. We conclude this because we found API-signatures that show injections into these processes, but not the remaining one. 

\begin{table}{}
\hspace*{-0.7cm} 
    \centering
    \footnotesize
    \begin{tabular}{l|c|c|c|c}
    
        & \multicolumn{4}{c}{\textbf{Precision}} \\
        & \multicolumn{4}{c}{(Minerva, PackerInspector)} \\
         Sample & Procs & Waves & API calls & Unique APIs\\
         \hline
        CryptoWall\footnote{md5sum e73806e3f41f61e7c7a364625cd58f65} & 5($\dagger$2),4($\dagger$1) & 8,4 & 7371,21050 & 148, 354 \\    
         CryptoWall \footnote{md5sum 5384f752e3a2b59fad9d0f143ce0215a} & 3, 4($\dagger$1) & 6,4 & 9945,23580 & 135, 388\\
         \hline
         Tinba \footnote{md5sum c141be7ef8a49c2e8bda5e4a856386ac} & 3,2 & 4,3 & 557,34076 & 54, 477 \\    
         Tinba \footnote{md5sum 08ab7f68c6b3a4a2a745cc244d41d213} & 3,2 & 4,3 & 667,49260 & 55,549 \\
         Tinba \footnote{md5sum 6244604b4fe75b652c05a217ac90eeac} & 3,2 & 4,3 & 704,49262 & 55, 550\\
         \hline
        Gapz \footnote{md5sum 089c5446291c9145ad8ac6c1cdfe4928} & 2,1 & 7,5 & 36509156, 15850000 & 140, 336 \\     
         Gapz \footnote{md5sum 0ed4a5e1b9b3e374f1f343250f527167}  & 2,1 & 4,3 & 36504908, 15845670 & 125, 226 \\
         Gapz \footnote{md5sum e5b9295e0b147501f47e2fcba93deb6c} & 3,1 & 5,2 & 36506063, 15844113 & 186, 251 \\
         \hline
         Ramnit \footnote{md5sum 448ce1c565c4378b310fa25b4ae3b17f} & 3, 4($\dagger$1) & 8,5 & 6908,56720 & 116, 479\\    
         Ramnit \footnote{md5sum 33cd65ebd943a41a3b65fa1ccfce067c} & 12($\dagger$1),5 & 30,6 & 16185,209828 & 153, 489 \\
         Ramnit \footnote{md5sum 3bb86e6920614ed9ac5d8fbf480eb437} & 3, 5($\dagger$1) & 8,8 & 3189, 115943 & 115, 621\\
    \end{tabular}
    \centering
    \caption{The evaluation results from matching Minerva and PackerInspector with the malware samples of data set \#2. $\dagger$ indicates the number processes we determined to be false positives.}
    \label{tab:malware_evaluation}
\end{table}

In terms of precision for tracking API calls, there is a similar relationship between Minerva and PackerInspector as when matched with our ground-truth samples. In the samples from CryptoWall, Tinba and Ramnit, Minerva reports roughly twelve times fewer API calls within the malware execution than PackerInspector. We manually investigated several of the Tinba samples to confirm these numbers and found that Minerva captures API calls accurately within the malware execution. As such, we consider the API calls reported by PackerInspector to be a significant over-approximation. In addition to the total number of API calls, PackerInspector captures about three times more unique API calls in each malware execution, even in cases where Minerva finds more total API calls. The only instances where Minerva finds more total API calls are in the Gapz samples. The reason that there is a significant amount of API calls in these cases is that the Gapz malware scans a remote process for gadgets and this results in an enormous amount of calls to \texttt{ReadProcessMemory} (about 99.985\% of calls in the Minerva analyses). We believe that the reason Minerva reports more API calls than PackerInspector only in the Gapz samples, is because PackerInspector exits prematurely. 

When matched with these malware samples, it is clear that PackerInspector over-approximates the API usage of the applications, both in terms of total API calls and unique APIs used. We also find that in the majority of times, PackerInspector under-approximates the system-wide malware propagation and Minerva finds more system-wide unpacking.

\begin{table}{}
    \centering
    
    \begin{tabular}{|l|c|c|c|c|}
    \hline
    1 & 2 & 3 & 4-5 & 6-11  \\
    \hline
    66\% & 15\% & 9\% & 5\%  & 5\%\\
    \hline
    \end{tabular}
    \caption{Number of process executions per malware sample.}
    \label{tab:proc_count}

    \begin{tabular}{|l|c|c|c|c|c|}
    \hline
    1 & 2 & 3 & 4 & 5 & 5 < \\
    \hline
    52\% & 18\% & 7\% & 9\% & 2\% & 12\%\\
    \hline
    \end{tabular}
    \caption{Number of waves per malware sample.}
    \label{tab:wave_count}

    \begin{tabular}{|l|c|c|c|c|c|}
    \hline
    1 & 2 & 3 & 4 & 5 & 5 < \\
    \hline
    51\% & 17\% & 8\% & 3\% & 8\% & 13\%  \\
    \hline
    \end{tabular}
    \caption{Number of PE files constructed per malware sample.}
    \label{tab:PE_count}

    \begin{tabular}{|l|c|c|c|}
    \hline
    1 & 2 & 3-5 & 5 < \\
    \hline
    56\% & 15\% & 18\% & 11\%  \\
    \hline
    \end{tabular}
    \caption{Number of Sections reconstructed per PE file.}
    \label{tab:section_count} 

    \begin{tabular}{|l|c|c|c|c|c|}
    \hline
    0 & 1-10 & 11-15 & 16-20 & 21-25 & 26 <   \\
    \hline
    18\% & 28\% & 7\% & 5\% & 2\% & 40\% \\
    \hline
    \end{tabular}
    \caption{Number of imports per PE file.}
    \label{tab:import_count}

\end{table}

\subsection{Relevance on malware}
In this experiment, we match Minerva with benchmark set \#4. In total we run 119 samples through Minerva and collect (1) the number of process executions; (2) the number of waves; (3) the number of generated PE files; (4) the number of imports in the IAT of each PE file and (5) the number of sections in each PE file. 

Table \ref{tab:proc_count} shows the number of processes and Table \ref{tab:wave_count} the number of waves in our data set. We find that a third of the samples perform multi-process execution and that roughly half have multiple execution waves, which means that a large part of all the samples with single-process execution have multi-wave execution. 

Table \ref{tab:PE_count} shows the distribution of reconstructed PE files. We construct more PE files than the number of captured waves, which shows that some waves contain several regions that are non-related. Finally, Table \ref{tab:section_count} shows the number of sections reconstructed in each PE file and Table \ref{tab:import_count} shows the number of reconstructed imports. For roughly 20\% of the PE files, we do not monitor any API calls in the code, and this is due to some PE files being a result of small amounts of taint in minor code regions.

\subsection{Relevance on packers}
In this experiment, we show that Minerva is relevant against publicly available packers from benchmark set \#3. This experiment is common practice for unpacking engines \cite{Bonfante:2015:CMS:2810103.2813627, 4413009, Royal:2006:PAH:1191820.1191885, Ugarte-pedrero_sok:deep} and, therefore, natural for us to perform. We construct a simple application that will get the name of the current user and report back to us so we can verify the behaviour occurred correctly. We pack this application with 13 publicly known packers and analyse the samples in Minerva.

We show the results of our experiment in Table \ref{tab:packer_relevance}. The table shows the number of processes, waves, PE files, and whether we found the original code or a derivative thereof, and also whether we observed the original behaviour. Minerva produced PE files for most of the packers that are very similar to the original code, including correct API calls. In general, these packers are rather simple in comparison to some of the techniques we observe in malware from the wild. For example, all but one of the packers are single-process packers. This makes sense since the packers are not necessarily meant to be used by malicious software, but may be used by benign applications, which are not meant to inject into other applications. Furthermore, many of these packers rely on similar approaches for compression, e.g. the Lempel–Ziv–Markov chain algorithm, and the majority the packers used in these experiments are rather old.

\begin{table}{}
    \centering
    
    \begin{tabular}{l|c|c|c|c|c}
    \hline
    Packer & \#proc & \#wave & \#PE & U & OB \\
    \hline
    BoxedApp & 1 & 1 & 1 & Y & Y \\
    Enigma & 2 & 11 & 1 & Y & Y \\
    FSG\_packed & 1 & 2 & 2 & Y & Y \\
    mew11 & 1 & 3 & 4 & Y & Y \\
    MoleBox & 1 & 4 & 9 & Y & Y \\
    mpress & 1 & 2 & 2 & Y & Y \\
    PackMan & 1 & 2 & 2 & Y & Y \\
    PECompact & 1 & 4 & 4 & Y & Y \\
    PEtite & 1 & 4 & 4 & Y & Y \\
    tElock & 0 & 0 & 0 & N & N \\
    UPX & 1 & 2 & 2 & Y & Y \\
    WinUpack & 1 & 2 & w & Y & Y \\
    XComp & 1 & 2 & 2 & Y & Y \\
    \end{tabular}
    \caption{The results from matching Minerva with known packers. \textbf{OB} indicates if we observed the original behaviour of the packed application. \textbf{U} indicates if we found the original code in Minerva's output.}
    \label{tab:packer_relevance}
\end{table}

\subsection{Tinba case study}
We now investigate in depth a case study of a real-world malware sample from the Tinba malware family\footnote{md5sum 08ab7f68c6b3a4a2a745cc244d41d213}. Minerva outputs four PE files with sizes 12KB, 12KB, 16KB and 24KB, respectively. We manually reverse engineered the sample to fully understand the system-wide propagation and where the sample exposes its unpacked code. The malware first decrypts memory from its data section and then transfers execution to this code. The decrypted code injects code into the Windows process \texttt{Winver.exe} and from \texttt{Winver.exe} it further injects into \texttt{explorer.exe}. 

To inject code into \texttt{Winver.exe} Tinba launches a new instance of \texttt{Winver.exe} in a suspended state. Then, Tinba allocates memory on the heap of the newly started \texttt{Winver.exe} and copies some malicious code into this specific memory. Tinba then overwrites six bytes of the \texttt{Start} function in \texttt{Winver.exe} with the instructions \texttt{push ADDR; ret}, where \texttt{ADDR} is some address inside the dynamically generated malicious code. Effectively, Tinba ensures execution of its malicious code in \texttt{Winver.exe} by overwriting an initial function in \texttt{Winver.exe} to hijack execution.

Minerva captures one execution wave and outputs two unpacked PE files for the code in \texttt{Winver\-.exe}, one PE file based on a single execution wave in \texttt{explorer.exe} and also one PE file from the unpacked code in the initial process. The PE files produced by Minerva has 0, 11, 12 and 26 imports reconstructed. These results capture the execution perfectly because the PE file with 0 imports is purely the \texttt{push ADDR; ret} instructions of the malware execution trace in \texttt{Winver\-.exe} and the rest of the PE files contain various other stages with more payload content. 

Minerva correctly identifies the malicious code, both the patched code of \texttt{Winver\-.exe} and also the code on the heap that contains the core of the malware code. More importantly, the PE files precisely capture the malware execution, and from the execution trace output by Minerva, we can precisely see the exact instructions \texttt{push ADDR; ret}. Minerva also catches the exact malware code inside of the \texttt{explorer.exe} process. The PE file captured from the second execution wave in the original malware process contains 11 imports in its reconstructed IAT, five of which are \texttt{ResumeThread, CreateProcessA, WriteProcessMemory, VirtualAllocEx} and  \texttt{VirtualProtectEx}. A novice analysts can quickly determine that the execution performs a code-injection based on these API calls.

\subsection{Performance evaluation}
\label{chapter4:PerformanceEvaluation}
In the final part of our evaluation, we monitor the performance of Minerva. The authors of PANDA report that recording gives a 1.85x slowdown in comparison to QEMU alone and replaying incurs a 3.57x slowdown \cite{Dolan-Gavitt:2015:RRE:2843859.2843867}. This is expensive in comparison to systems that rely on hypervisor-based virtualisation for recording, e.g. AfterSight \cite{Chow:2008:DDP:1404014.1404015}. However, we consider PANDA's performance good enough for malware analysis in particular because the plugins that we deploy will have far more impact on the total analysis time. Naturally, the performance overhead in the recording stage can be used by the malware to evade analysis, and we discuss this further in Section \ref{Chapter4:limitations}. In this performance evaluation, we focus on the overhead of Minerva's analysis when replaying the recorded execution, and the numbers we report in this section are based on analysis of the 25 malware samples from the Ramnit, Gapz, CryptoWall and Tinba families.

\begin{figure}[H]
\centering
\label{ScoreVersusSize}
    \begin{tikzpicture}
        \begin{axis}[
        scale=0.55,        
            width=0.8\textwidth,
    height=0.4\textwidth,
        xmin=0, xmax=1725000000,
    ymin=0, ymax=7000,
                xlabel=Instructions replayed,
                ylabel=Seconds]
            \addplot+[error bars/.cd,
                       y dir=both, y explicit]
                    coordinates {
(75000000,283) +- (42, 42)
(150000000,512) +- (65, 65)
(225000000,704) +- (96, 96)
(300000000,850) +- (114, 114)
(375000000,995) +- (146, 146)
(450000000,1121) +- (147, 147)
(525000000,1274) +- (157, 157)
(600000000,1475) +- (174, 174)
(675000000,1700) +- (197, 197)
(750000000,1893) +- (218, 218)
(825000000,2128) +- (253, 253)
(900000000,2324) +- (239, 239)
(975000000,2616) +- (243, 243)
(1050000000,2805) +- (266, 266)
(1125000000,3013) +- (320, 320)
(1200000000,3326) +- (393, 393)
(1275000000,3359) +- (457, 457)
(1350000000,3583) +- (482, 482)
(1425000000,3661) +- (556, 556)
(1500000000,3902) +- (543, 543)
(1575000000,4124) +- (610, 610)
(1650000000,4207) +- (706, 706)
(1725000000,3973) +- (680, 680)
                    };
                \addlegendentry{LLVM + taint + Minerva}
                \addplot[error bars/.cd,
                       y dir=both, y explicit] coordinates {
(75000000,213) +- (54, 54)
(150000000,265) +- (42, 42)
(225000000,350) +- (58, 58)
(300000000,406) +- (65, 65)
(375000000,441) +- (66, 66)
(450000000,489) +- (57, 57)
(525000000,537) +- (56, 56)
(600000000,606) +- (57, 57)
(675000000,679) +- (75, 75)
(750000000,714) +- (78, 78)
(825000000,761) +- (83, 83)
(900000000,854) +- (92, 92)
(975000000,964) +- (117, 117)
(1050000000,1030) +- (137, 137)
(1125000000,1089) +- (155, 155)
(1200000000,1158) +- (155, 155)
(1275000000,1043) +- (111, 111)
(1350000000,1031) +- (82, 82)
(1425000000,1005) +- (87, 87)
(1500000000,1099) +- (111, 111)
(1575000000,1163) +- (118, 118)
(1650000000,1110) +- (103, 103)
(1725000000,1106) +- (96, 96)
                    };
                    \addlegendentry{LLVM + taint}
        \end{axis}
    \end{tikzpicture}
    \caption{The average number and standard deviation of instructions replayed relative to time, for instruction counts where we have more than 3 samples executing the given number of instructions. }
\label{fig:InstructionsRelativeToTime}    
\end{figure}
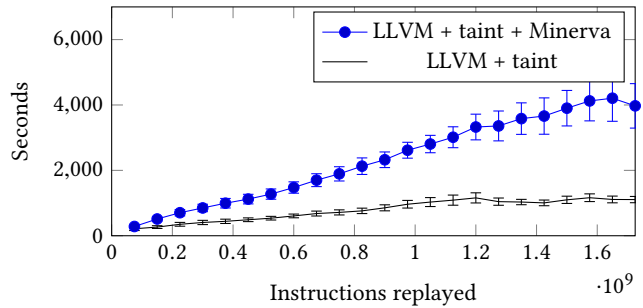

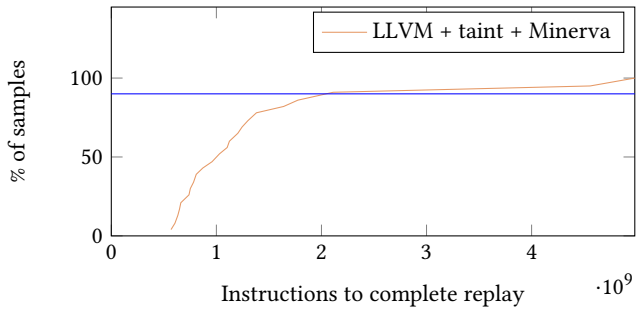
\begin{figure}[H]
\centering
\label{ScoreVersusSize}
    \begin{tikzpicture}
        \begin{axis}[
        scale=0.55,
            width=0.8\textwidth,
    height=0.4\textwidth,
        xmin=0, xmax=4982041422,
    ymin=0, ymax=145,
                xlabel=Instructions to complete replay,
                ylabel=\% of samples]
            \addplot+[mark = none, red]
                    coordinates {
(568380645, 4)
(605545483, 8)
(633190990, 13)
(649319669, 17)
(661589381, 21)
(739277161, 26)
(752206384, 30)
(782672024, 34)
(808381032, 39)
(870838726, 43)
(958901927, 47)
(1029782762, 52)
(1102900266, 56)
(1124087343, 60)
(1204063468, 65)
(1244763771, 69)
(1299911044, 73)
(1381355627, 78)
(1640070664, 82)
(1775595713, 86)
(2112022315, 91)
(4557122506, 95)
(4982041422, 100)
                    };
                \addlegendentry{LLVM + taint + Minerva}
            \addplot+[mark = none, blue]
                    coordinates {
(0, 90)
(4982041422, 90)
                    };
                             
        \end{axis}
    \end{tikzpicture}
    \caption{The amount of instructions needed to replay the samples in our data set. The horizontal blue line shows the 90\% mark. }
    \label{chapter5:figInstructionSampleCount}
\end{figure}

The blue curve of Figure \ref{fig:InstructionsRelativeToTime} shows the number of instructions replayed relative to the time taken in each of the analyses. On average the replay time is 3166 seconds, resulting in a 126x slowdown of the recording time, and we analysed on average 432365 instructions per second. In comparison, the developers of PANDA report a 24.7x slowdown when tainting data sent over the network and a 67.7x slowdown for tainting a 1KB file and encrypting it with AES-CBC-128 \cite{Dolan-Gavitt:2015:RRE:2843859.2843867}. Additionally, Figure \ref{chapter5:figInstructionSampleCount} shows the number of instructions that it took to replay the samples in our data set, and we observe that for about 90\% of the samples this required less than 2 billion instructions. 

Another interesting metric is the specific overhead incurred by Minerva-only code. Specifically, there is some share of the overhead that is due to the translation of QEMU TCG instructions to LLVM instructions and also overhead that is specific to the taint implementation of PANDA. None of these requirements is strict to Minerva, in that we are not reliant on LLVM specifically, and PANDA's taint analysis does not focus on performance.  Several systems focus on fast taint analysis \cite{argos:eurosys06, Bosman:2011:MWF:2186328.2186330, Henderson:2017:DPW:3057931.3057958} and, conceptually, the techniques of Minerva can be implemented on top of these taint libraries as well. To understand the overhead of Minerva's code, we ran the samples through a replay with LLVM-translation and taint analysis enabled, and no Minerva-specific analysis code. This gives us a reasonable estimate for how much of the analysis time was spent in the specific code related to Minerva. The black curve in Figure \ref{fig:InstructionsRelativeToTime} shows these numbers. On average, each execution took 1275 seconds with the overhead of LLVM translation and PANDA's taint library. This corresponds to an average of 51x slowdown, meaning that Minerva's code takes up a bit more than half of the total 126x slowdown.

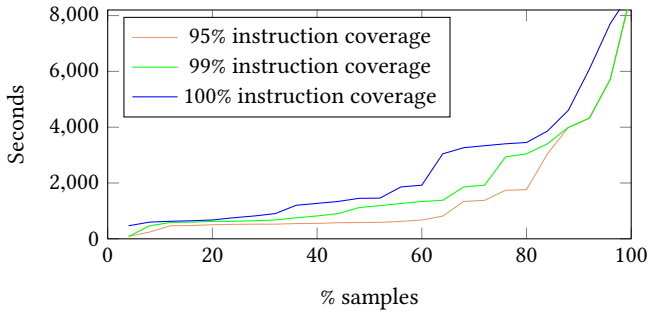
\begin{figure}[H]
\centering
\begin{tikzpicture}
\begin{axis}[
        scale=0.55,
    width=0.8\textwidth,
    height=0.4\textwidth,
    xlabel={\% samples},
    ylabel={Seconds},
    scaled ticks=false,
    ylabel shift = 1 pt,
    xmin=0, xmax=100,
    ymin=0, ymax=8200,
    legend pos=north west,]
\addplot[
    color=red,]
    coordinates {
(4.000000, 84.760000)
(8.000000, 242.000000)
(12.000000, 467.950000)
(16.000000, 476.270000)
(20.000000, 504.800000)
(24.000000, 517.380000)
(28.000000, 524.590000)
(32.000000, 525.260000)
(36.000000, 546.180000)
(40.000000, 551.330000)
(44.000000, 577.070000)
(48.000000, 584.700000)
(52.000000, 590.310000)
(56.000000, 624.040000)
(60.000000, 675.750000)
(64.000000, 812.970000)
(68.000000, 1338.220000)
(72.000000, 1379.300000)
(76.000000, 1738.510000)
(80.000000, 1765.380000)
(84.000000, 3045.460000)
(88.000000, 3991.120000)
(92.000000, 4321.260000)
(96.000000, 5716.470000)
(100.000000, 8751.300000)
    };

\addplot[
    color=green]
    coordinates {
    (4.000000, 84.760000)
(8.000000, 467.950000)
(12.000000, 584.700000)
(16.000000, 597.780000)
(20.000000, 624.040000)
(24.000000, 628.710000)
(28.000000, 646.370000)
(32.000000, 675.750000)
(36.000000, 753.580000)
(40.000000, 818.460000)
(44.000000, 903.130000)
(48.000000, 1118.000000)
(52.000000, 1189.790000)
(56.000000, 1265.180000)
(60.000000, 1338.220000)
(64.000000, 1379.300000)
(68.000000, 1857.970000)
(72.000000, 1922.340000)
(76.000000, 2934.180000)
(80.000000, 3045.460000)
(84.000000, 3405.290000)
(88.000000, 3991.120000)
(92.000000, 4321.260000)
(96.000000, 5716.470000)
(100.000000, 8846.860000)};

\addplot[
    color=blue]
    coordinates {
(4.000000, 467.950000)
(8.000000, 597.780000)
(12.000000, 628.710000)
(16.000000, 646.370000)
(20.000000, 675.750000)
(24.000000, 753.580000)
(28.000000, 818.460000)
(32.000000, 903.130000)
(36.000000, 1201.710000)
(40.000000, 1269.260000)
(44.000000, 1338.220000)
(48.000000, 1450.370000)
(52.000000, 1458.120000)
(56.000000, 1857.970000)
(60.000000, 1922.340000)
(64.000000, 3045.460000)
(68.000000, 3264.350000)
(72.000000, 3336.690000)
(76.000000, 3405.290000)
(80.000000, 3451.720000)
(84.000000, 3860.760000)
(88.000000, 4607.140000)
(92.000000, 6077.810000)
(96.000000, 7711.970000)
(100.000000, 8846.860000)
  };
\legend{95\% instruction coverage, 99\% instruction coverage, 100\% instruction coverage}
    \end{axis}
\end{tikzpicture}
\caption{The time taken to explore the unique instructions in each malware sample.}
\label{fig:CFGInstrsCoverage}
\end{figure}

During the replay of a malware sample, Minerva does no checking to verify if the analysis is progressing, is stuck or something similar, and the numbers above report the total time of each analysis-replay. An interesting metric in addition to the total replay time is the time it took to reveal the instructions executed in each malware sample. In Figure \ref{fig:CFGInstrsCoverage} we show the time it took to uncover 95\%, \%99 and \%100 of the unique instructions executed by the malware samples, respectively. The numbers decrease significantly, and on average it took 1614, 1964 and 2543 seconds to uncover 95\%, \%99 and \%100 of the unique instructions executed, respectively. As such, it took roughly half of the total replay time to reveal 95\% of the instructions in each sample. 

\begin{figure}[H]
\centering
\includegraphics[scale=0.45]{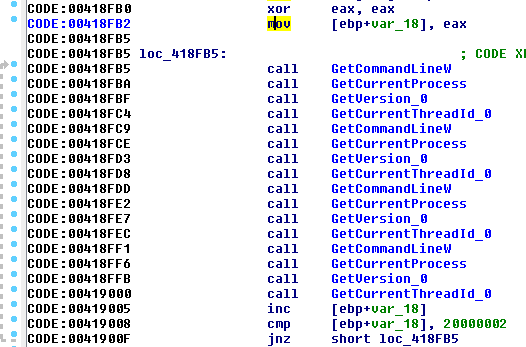}
\caption{Stalling loop in Kovter malware.}
\label{fig:StallingLoopKovterSample5cc6f7}
\end{figure}

The biggest performance bottleneck we found in Minerva is when malware makes the code execute longer via stalling loops. An example of a stalling loop from a Kovter sample\footnote{md5 of sample 147330a7ec2e27e2ed0fe0e921d45087} is shown in Figure \ref{fig:StallingLoopKovterSample5cc6f7}. In total, the loop does 20 million iterations with sixteen calls to functions from the Windows API in each iteration. The loop has no real effect and is purely garbage code. In total, our 25-second recording of this sample reaches 17 million iterations before the recording is over and incurs a replay time of 3300 seconds. 

The sample we observed with the longest replay time is from the Nymaim family that has a stalling loop with 1.4 billion iterations, and after the stalling loop, it calls the \texttt{Sleep} function from the Windows API to further stall the execution. In total, our 25-second recording of this sample took 170,000 seconds to replay. 

\section{Limitations}
\label{Chapter4:limitations}
\textbf{Stolen bytes and copying Windows API.} Minerva's precise API capturing of API calls depends on monitoring whether the target address of branch instructions is the start of some Windows API function. Some malware use an anti-analysis technique, called \textit{stolen bytes}, that copies a share of some API function to another place in memory to execute that code and then branch in the middle of the given API function. In this context, they avoid calling the beginning of the function and our technique will not capture it. One solution to this is to identify function boundaries for each API function and then monitor ranges rather than the function start. 

In a more general setting, malware can copy entire functions or modules from the Windows API and then rely on the copied code rather than calling the original Windows code. In this context, Minerva will still capture whenever system calls happen, but new measurements should be taken for identifying the copying of Windows code. One approach is to mark library code with a specific taint label and then monitor whether library code is propagated. Naturally, this solution is subject to the limitations of taint analysis. Another approach is to incorporate forensic techniques that determine the similarity between the code in a given process and a set of external libraries, which we discuss further in the following paragraph.\\

\textbf{Inlining and statically linked binaries.} A limitation in Minerva in terms of identifying external dependencies is when malware deploys inlined or statically linked code. The difference between this and copying external libraries as described above is that inlining and statically linking occurs at compile time where copying occurs at run time. Minerva is not capable of identifying inlined or statically linked external dependencies, and we consider this to be a slightly different problem, namely similarity analysis of the malware code with library implementations. However, inlining and statically linking can, of course, be used in combination with obfuscation techniques and similar, and, therefore, the problem becomes determining program equivalence in the general case, which is a well-known undecidable problem. Nonetheless, efforts can still yield positive and practical results as shown by previous work in areas such as library fingerprinting \cite{Emmerik1994SignaturesFL, Jacobson:2011:LLF:2024569.2024571}, structural comparison of binary code \cite{articleDullienAndRolf, DBLP:conf/dimva/Flake04, Kruegel:2005:PWD:2146257.2146273} and, most recently, similarity detection via machine learning \cite{li2019graph, DBLP:conf/ccs/SongYLS18}.  \\

\textbf{Performance limitations.} There are currently two main performance limitations in Minerva. First, malware can detect the presence of the recording component due to the 3.56x slowdown, and second, the replaying component limits the throughput of Minerva due to its performance cost. Stalling loops seem to pose a core limitation in this context. There is, however, previous work on how to deal with stalling loops in the context of full system emulation. Kolbitsch et al. implemented several features into the Anubis analysis system \cite{Kolbitsch:2011:PPD:2046707.2046740}. Their approach is to implement heuristics that detect when stalling loops occur and then either disable heavy instrumentation until the stalling loop exits or force execution out of the loop. The first approach is certainly possible to implement in Minerva, but it may run into issues if the stalling loop is also responsible for propagating executable malicious code since the taint analysis would likely be disabled. The second approach is more challenging to implement because replaying is not able to change execution-path in the guest system, as the execution is fixed to the replay log.

We consider five main avenues to improving performance. First, we can use hardware assisted virtualisation during recording and only full system emulation during replay, as suggested in AfterSight \cite{Chow:2008:DDP:1404014.1404015}. Second, we can implement various on-and-off analyses during the replay similar to Kolbitsch et al. Third, we can add light anti-analysis monitoring during the recording, for example, to limit the effectiveness of calls to functions like \texttt{Sleep}. In this case however, the implementation must use some form of approximation to determine the malware execution trace since taint analysis will not be available. Fourth, we can improve the speed of various parts in PANDA, such as the taint analysis plugin. Instead of converting instructions to LLVM and performing taint analysis on the LLVM code, we can adopt the taint system by DECAF, which occurs directly on the QEMU tcg instructions \cite{Henderson:2017:DPW:3057931.3057958}. Finally, an interesting avenue is implementing a feedback loop between record-and-replay that based on the analysis in the replay sends information to the recording about where a delay in execution occur and how to handle it. In this way, it is possible to incrementally build up a complete execution trace of the malware without anti-analysis tricks.

\section{Related work}
\label{sec:Chapter4RelatedWork}
\textbf{Automatic unpacking.} There are many works in automatic unpacking of malware and we have already discussed several of these throughout the paper \cite{Dinaburg:2008:EMA:1455770.1455779, Hu:2013:MSM:2535461.2535485, Josse2007, Kang:2007:RHC:1314389.1314399,  4413009, 10.1007/978-3-540-88313-5_31,  Ugarte-pedrero_sok:deep}. Some of this work considers the concept of IAT destruction \cite{Josse2007, 10.1007/978-3-540-88313-5_31, DBLP:conf/malware/Korczynski16} and IAT reconstruction has also been considered on a more general basis \cite{DBLP:journals/jip/KawakoyaIM18}. The work by Ugarte et al. \cite{Ugarte-pedrero_sok:deep} highlights several missing gaps in existing unpackers and proposes a system-wide approach to unpacking. However, as we observed in this paper, their approach is severely limited. In some aspects, Ugarte et al. provide a more refined model for dynamically generated code in that they assign various labels to the memory written by the malware based on whether it is executed and alike. These labels can easily be integrated into Minerva. In addition to this, they also highlight that several limitations in existing unpackers exist due to missing reference data sets, which indeed also motivated the construction of our synthetic benchmark set \#1.

The work that is closest to ours is Tartarus \cite{DKOR} and the ideas of this paper are heavily inspired by their work. We deploy a similar approach to tracing the malware throughout the whole system, however, we deploy a different model of dynamically generated malicious code and also propose novel algorithms for making the output suitable for follow-up analysis. In particular, the post-processing we describe in this paper is novel and our model of dynamically generated code is explicitly connected to previous waves whereas Tartarus simply dumps the whole of tainted memory whenever a new wave executions. As such, our model is more precise and also formally defined. \\

\textbf{System-wide malware execution.} Several works have closely considered the concept of malware executing throughout the whole system. In particular, Panorama \cite{Yin:2007:PCS:1315245.1315261}, DiskDuster\cite {10.1007/978-3-642-37300-8_9}, Tartarus \cite{DKOR} and API Chaser \cite{YuheiKawakoya2019} use dynamic taint analysis to capture this. Barabosch et al. has also investigated the problem with code injection by analysing memory dumps \cite{10.1007/978-3-319-60876-1_10} and also at run time \cite{10.1007/978-3-319-08509-8_13}. Minerva relies on the same techniques as Tartarus to trace malware execution through the system. An interesting approach at the other end of the spectrum is explored by Ispoglou and Payer in malWASH \cite{198415}, where they propose to write complex malware using exactly the paradigm of system-wide execution.  \\

\textbf{Malware disassembly.} The work in this paper is closely related to techniques that focus on disassembling malicious software. An accurate description of our work within this domain, rather than unpacking, is a system-wide malware disassembler. We gather a precise instruction-level execution trace of the malware and then gather more content to include in the reconstructed PE file with speculative disassembly. Traditionally, disassembly techniques are split between linear sweep, as used in GNU's Objdump, and recursive traversal \cite{Cifuentes:1995:DBP:213593.213604, Sites:1993:BT:151220.151227} algorithms. However, there are several pieces of previous work on disassembly that specifically target malware and these move beyond the traditional approaches. Kruegel et al. present an approach that combines a variety of techniques from control-flow analysis and statistical methods, in order to statically disassemble obfuscated binaries \cite{Kruegel:2004:SDO:1251375.1251393}. Kinder and Veith present an approach based on abstract interpretation that statically disassembles binaries and also resolves indirect branch instructions \cite{Kinder:2008:JSA:1427782.1427835}. Rosenblum et al. present a classification approach to identify function entry points \cite{DBLP:conf/aaai/RosenblumZMH08}, and Bao et al. \cite{DBLP:conf/uss/BaoBWTB14} follow the same path and use machine learning and static analysis to identify functions within binaries. They train a weighted-prefix tree that recognises function starting points in a binary file and then value-set analysis \cite{Balakrishnan2008} with an incremental control-flow recovery algorithm to identify function boundaries.

\section{Conclusions}
In this paper, we proposed a system called Minerva that focuses on generic and precise malware unpacking. From a technical point of view, Minerva deploys a concatic approach with both dynamic and static analysis and partitions the malware execution trace into execution waves based on information flow analysis. Minerva precisely monitors the API-calls of the malware code and accurately correlates these to the unpacked code. Based on the output of the dynamic analysis, Minerva performs static analysis on the execution waves to output a set of reconstructed PE files with valid import address tables and patched API calls. 

From a theoretical point of view, Minerva deploys a precise model of execution waves based on an information flow model that captures dynamically generated malicious code \textit{independently} of who wrote the code. We came up with several novel algorithms that combine these execution waves with other artefacts collected from the dynamic analysis to carefully produce PE files that are well-suited for follow-up static analysis. 

Finally, we proposed a new set of benchmark applications that exhibit unpacking behaviours with various forms of dynamically generated code, system-wide execution and import-address table destruction in order to address a missing gap in terms of ground-truth samples for testing unpackers. This benchmark suite is the first of its kind in that previous benchmark data sets for testing automatic unpackers rely on third-party applications to perform the packing. 

We evaluated Minerva against our synthetic applications, real-world malware samples and also performed a comparative evaluation. Our results show that Minerva is significantly more precise than previous work and outputs unpacked code that shows external dependencies, which previous work does not. Our results also show that Minerva captures system-wide unpacking in many cases where previous work fails. 

\bibliographystyle{ACM-Reference-Format}
\bibliography{sample-sigconf}

\end{document}

%% file: AlgorithmStyle.tex
\usepackage{algorithm2e}
\usepackage{xcolor}

\SetAlFnt{\footnotesize}

\SetAlCapSty{xAlCapSty}

\SetCommentSty{xCommentSty}

\SetNlSty{mynlfont}{}{} 

\LinesNumbered

\SetSideCommentRight

\DontPrintSemicolon

\RestyleAlgo{algoruled}

\definecolor{mGreen}{rgb}{0,0.6,0}
\definecolor{mGray}{rgb}{0.5,0.5,0.5}
\definecolor{mPurple}{rgb}{0.58,0,0.82}
\definecolor{backgroundColour}{rgb}{0.95,0.95,0.92}

\lstdefinestyle{CStyle}{
    backgroundcolor=\color{backgroundColour},   
    commentstyle=\color{mGreen},
    keywordstyle=\color{magenta},
    numberstyle=\tiny\color{mGray},
    stringstyle=\color{mPurple},
    basicstyle=\footnotesize,
    breakatwhitespace=false,         
    breaklines=true,                 
    captionpos=b,                    
    keepspaces=true,                 
    numbers=left,                    
    numbersep=5pt,                  
    showspaces=false,                
    showstringspaces=false,
    showtabs=false,                  
    tabsize=2,
    language=C
}

%% file: mytikz.tex
\usepackage[utf8]{inputenc}

\usepackage{tikz}
\usetikzlibrary{shapes,shadows}
\usetikzlibrary{arrows,automata}
\usetikzlibrary{positioning}
\usetikzlibrary{calc,positioning}
\usetikzlibrary{decorations.markings}

\makeatletter
\tikzset{%
  remember picture with id/.style={%
    remember picture,
    overlay,
    save picture id=#1,
  },
  save picture id/.code={%
    \edef\pgf@temp{#1}%
    \immediate\write\pgfutil@auxout{%
      \noexpand\savepointas{\pgf@temp}{\pgfpictureid}}%
  },
  if picture id/.code args={#1#2#3}{%
    \@ifundefined{save@pt@#1}{%
      \pgfkeysalso{#3}%
    }{
      \pgfkeysalso{#2}%
    }
  }
}

\def\savepointas#1#2{%
  \expandafter\gdef\csname save@pt@#1\endcsname{#2}%
}

\def\tmk@labeldef#1,#2\@nil{%
  \def\tmk@label{#1}%
  \def\tmk@def{#2}%
}

\tikzdeclarecoordinatesystem{pic}{%
  \pgfutil@in@,{#1}%
  \ifpgfutil@in@%
    \tmk@labeldef#1\@nil
  \else
    \tmk@labeldef#1,(0pt,0pt)\@nil
  \fi
  \@ifundefined{save@pt@\tmk@label}{%
    \tikz@scan@one@point\pgfutil@firstofone\tmk@def
  }{%
  \pgfsys@getposition{\csname save@pt@\tmk@label\endcsname}\save@orig@pic%
  \pgfsys@getposition{\pgfpictureid}\save@this@pic%
  \pgf@process{\pgfpointorigin\save@this@pic}%
  \pgf@xa=\pgf@x
  \pgf@ya=\pgf@y
  \pgf@process{\pgfpointorigin\save@orig@pic}%
  \advance\pgf@x by -\pgf@xa
  \advance\pgf@y by -\pgf@ya
  }%
}

\makeatother

\usepackage{caption}

\usepackage{adjustbox}

\usetikzlibrary{positioning}
\usetikzlibrary{fit,backgrounds}

\tikzset{
    state/.style={
           rectangle,
           rounded corners,
           draw=black, very thick,
           minimum height=2em,
           inner sep=2pt,
           },
    propstate/.style={
           rectangle,
           draw=black, very thick,
           minimum height=2em,
           inner sep=2pt,
           },
}

\usepackage{color, colortbl}
\definecolor{Gray}{gray}{0.9}
\definecolor{lightblue}{rgb}{0.93,0.95,1.0}
\definecolor{myyellow}{rgb}{0.95, 0.95, 0}
\definecolor{navy_blue}{rgb}{0.7, 0.7, 0.90}
\definecolor{red}{rgb}{0.9, 0.6, 0.4}
\definecolor{pale_green}{rgb}{0.75, 0.96, 0.75}

\pgfdeclarelayer{background}
\pgfdeclarelayer{foreground}
\pgfsetlayers{background,main,foreground}

\tikzstyle{plain}=[draw, fill=gray!00, text width=10em, text centered, minimum height=1.5em]
\tikzstyle{thickplain}=[draw, fill=gray!00, text width=5em, text centered, minimum height=2.5em]
\tikzstyle{malplain}=[draw, fill=gray!00, text width=4em, text centered, minimum height=2.5em]

\tikzstyle{textplain}=[text width=10em, text centered]

\tikzstyle{addressbox}=[draw, fill=gray!00, text width=5em, text centered, minimum height=1.5em]

\tikzstyle{sensor}=[draw, fill=blue!20, text width=5em, 
    text centered, minimum height=2.5em,drop shadow]
\tikzstyle{ann} = [above, text width=5em, text centered]
\tikzstyle{wa} = [sensor, text width=10em, fill=red!20, 
    minimum height=6em, rounded corners, drop shadow]
\tikzstyle{sc} = [sensor, text width=13em, fill=red!20, 
    minimum height=10em, rounded corners, drop shadow]
\tikzstyle{wave} = [draw, fill=gray!30, text width=4em, 
    text centered, minimum height=2.5em,drop shadow, rounded corners, drop shadow]
\tikzstyle{wave2} = [draw, fill=gray!30, text width=7em, 
    text centered, minimum height=2.5em,drop shadow, rounded corners, drop shadow]

\tikzstyle{code} = [draw, fill=gray!20, 
    minimum height=2.5em,drop shadow]
\tikzstyle{codereuse} = [draw, fill=black!75, text width = 5em,
	text centered, minimum height=2.5em, drop shadow,text=white]

\tikzstyle{docs} = [draw, minimum height=4em, minimum width=3em, 
                fill=white, 
                double copy shadow={shadow xshift=-4pt, 
                             shadow yshift=-4pt, fill=white, draw}]